\documentclass[preprint,10pt]{elsarticle} \usepackage{lineno} \usepackage{amsmath,amssymb} \usepackage{graphicx} \usepackage{hyperref} \usepackage{makecell} \graphicspath{{figures/}{paper5/figures/}{../paper5/figures/}}

\newcommand{\MP}{M_P}
\newcommand{\MU}{M_U}
\newcommand{\Cinfl}{C_{\rm infl}}
\newcommand{\Neff}{N_{\rm eff}}
\newcommand{\OmL}{\Omega_\Lambda}
\newcommand{\Omm}{\Omega_m}

\journal{New Astronomy}

\begin{document}

\begin{frontmatter}

\author{Martin Drobczyk}
\ead{martin.drobczyk@dlr.de} \affiliation{organisation={German Aerospace Center (DLR e.V.)}, addressline={Robert-Hooke-Str. 7}, postcode={28359}, city={Bremen}, country={Germany}}

\title{A cross-epoch endpoint-consistency test of a single effective scaling from dark energy to inflation} 

\begin{abstract}
A cross-epoch endpoint-consistency test is formulated for a single power-law effective scaling, $M_U(\mu)=\beta^{1/4}\,M_P(\mu/M_P)^\gamma$, connecting late-time cosmic acceleration to the inflationary energy scale. The normalization is anchored at $\mu=H_0$ by the late-time dark-energy closure relations of the density-responsive gravity (DRG) framework and extrapolated to the inflationary Hubble rate $\mu=H_*$. Comparison with the CMB-normalized Starobinsky plateau defines the residual matching factor $C_{\rm infl}\equiv V_0^{A_s}/V_0^{\rm RG}$. The main novelty is not the generic idea that inflation and dark energy may be related, but the formulation of that question as a quantitative endpoint test between two empirically anchored energy scales. More generally, any framework supplying a late-time anchor scale is subject to the same hierarchy-enhanced endpoint filter, because the lever arm $H_*/H_0\sim 10^{55}$ compresses endpoint uncertainties into $\delta\gamma\propto [p\ln(H_*/H_0)]^{-1}\delta\ln X$. Requiring $C_{\rm infl}=\mathcal{O}(1)$ therefore selects a narrow consistency band. For the benchmark operator scaling $V_0\propto M_U^4$ with $c_4=\mathcal{O}(1)$, the matching requires $\gamma\simeq 0.491$ and yields $\beta\simeq 0.68$, both genuinely $\mathcal{O}(1)$. Varying the operator dimension over $p\in[2,8]$ shifts the required $\gamma$ only mildly (approximately $0.45$--$0.52$), and natural endpoint matching favors $p\simeq 3$--$4$. We present the full uncertainty budget, including cumulative threshold effects parametrized by $\Xi$ and observational errors on the late-time anchor, parametrized here through $w_0$, together with $A_s$ and $H_*$. Within the DRG benchmark, the primary scaling-sensitive result is therefore the narrow endpoint-selected window in $\gamma$ rather than the benchmark value of $w_0$ by itself. A secondary consequence is that the effective vacuum sector dilutes as $\rho_\Phi\propto a^{-6\gamma}$ at late times, faster than spatial curvature for $\gamma>1/3$. In a closed universe, this permits a cosmological turnaround at curvature values of order $|\Omega_K|\sim\mathcal{O}(10^{-4})$, testable with Stage-IV surveys and qualitatively distinct from $\Lambda$CDM.
\end{abstract}



\end{frontmatter}

\section{Introduction}
\label{sec:intro}

Late-time cosmic acceleration and primordial inflation define two vacuum-like energy scales separated by an enormous hierarchy. The present dark-energy scale, $\rho_\Lambda^{1/4}\sim\mathrm{few}\;\mathrm{meV}$, lies roughly 120 orders of magnitude below the Planck scale~\cite{Weinberg1989,Martin2012}, while the CMB-normalized inflationary plateau height $V_0^{1/4}\sim 10^{16}\,\mathrm{GeV}$ remains many orders below $M_P$ and about 110 orders above $\rho_\Lambda$. Standard cosmology treats these scales independently. Any framework that links them with a single scaling law therefore makes simultaneous statements about late-time dark energy and about the normalization of the inflationary plateau. In the benchmark construction adopted below, this directly affects the equation of state $w_0$, while tensor observables such as $r$ remain tied to the fixed inflationary potential shape.

Recent observations make this question more concrete. DESI~DR2 baryon acoustic oscillation (BAO) measurements, combined with CMB and supernova data, can prefer time-evolving dark energy at the few-$\sigma$ level, depending on the dataset combination~\cite{DESI2024,DESI_DR2_2025}. The Dark Energy Survey (DES) Y6 analysis finds $w_0=-0.981^{+0.021}_{-0.022}$ in a $w_0$CDM fit~\cite{DESY6_2024}. These data do not establish $w_0\neq -1$, but they leave room for percent-level departures from a cosmological constant, precisely the regime relevant for the present endpoint test.

Several approaches have been proposed to connect inflation and dark energy within a single theoretical structure. Quintessential inflation models~\cite{Peebles_Vilenkin1999,Dimopoulos_Owen2017} use a single scalar field whose potential interpolates between a slow-roll plateau and a late-time runaway, typically requiring specific potential shapes. Running vacuum models~\cite{Shapiro_Sola2009,SolaPeracaula2022} parametrize $\rho_\Lambda(H)$ as a series in $H^2$, connecting the vacuum energy to the expansion rate. Modified gravity approaches~\cite{Starobinsky2007,Nojiri_Odintsov2011} can produce both early and late acceleration through geometric terms, though they generally require model-specific Lagrangians. The present work differs from these programs in that it formulates the cross-epoch connection as an endpoint consistency condition between two empirically anchored energy scales, rather than as a dynamical evolution equation or a specific potential. The novelty therefore does not lie in the broad idea that inflation and dark energy might be connected. It lies in turning that idea into a quantitative benchmark test with a narrow viable scaling window, an explicit uncertainty budget, and concrete failure criteria. The endpoint-consistency test itself is not tied in principle to a unique late-time framework. What is required is a concrete late-time anchor that connects the observed dark-energy sector to an effective mass scale at $\mu=H_0$. Section~\ref{sec:universal_filter} derives the general endpoint filter in that model-independent form before specializing to a benchmark realization.

In this paper that benchmark is the density-responsive gravity (DRG) framework~\cite{Drobczyk2025DRG}. DRG supplies the needed late-time anchor through explicit density-responsive scalar-field closure relations that tie together $A$, $w_0$, $s_0$, and $\MU^4(H_0)$ (Section~\ref{sec:rho_phi}). In the fiducial mode used below, we adopt $A=0.024$ as a benchmark assumption together with the observed Planck 2018 background densities, which implies $w_0=-0.989$, $s_0=0.0112$, and $\MU^4(H_0)$. In the sensitivity scans, we instead vary $w_0$ within observational bounds and derive the corresponding $A(w_0)$ from the same closure relations. The detailed DRG construction is reviewed in Section~\ref{sec:framework}. Here the key point is simply that DRG makes the endpoint test calculable. We also stress that DRG is not a MOND-like low-acceleration framework. It does not alter the Newtonian force law at galactic scales, but operates in the cosmological dark-energy sector.

The central hypothesis is that the mass scale $M_U$ runs with an effective power law,
\begin{equation}
\MU(\mu) = \beta^{1/4}\,\MP\left(\frac{\mu}{\MP}\right)^\gamma,
\label{eq:MU_intro}
\end{equation}
where $\MU$ denotes the running characteristic mass scale of the density-responsive scalar sector~\cite{Drobczyk2025DRG}, and $\mu$ is the renormalization scale, identified with the Hubble rate $H$ in quasi-de~Sitter regimes. Detailed motivation and alternative identifications are discussed in Section~\ref{sec:RG_scale} and \ref{app:scale_setting}. Here $\MP\equiv (8\pi G)^{-1/2}\simeq 2.435\times 10^{18}\,\mathrm{GeV}$ is the reduced Planck mass, $\beta$ is a dimensionless normalization constant encapsulating threshold corrections, scheme dependence, and possible frame-transformation factors, and $\gamma$ is the effective scaling exponent. The exponent $\gamma\sim 0.5$ is motivated by anomalous dimensions in near-conformal gauge theories~\cite{Appelquist1988,Sannino2004,DeGrand2015}, but the present analysis treats it as a free parameter to be determined by consistency. Late-time cosmology fixes the product $\beta\,(H_0/\MP)^{4\gamma}$ through the density-responsive scalar-field closure relations (derived quantity, not assumed). The inflationary plateau height, fixed by the observed CMB amplitude $A_s$ (observational anchor), provides a second condition. Together these two anchors determine $\gamma$ algebraically, because the lever arm $H_*/H_0\sim 10^{55}$ makes the matching exponentially sensitive to the exponent.

The present study is an effective-theory benchmark analysis, not a derivation from a fundamental UV completion. We do not claim that inflation and dark energy are driven by the same dynamical field. We do not derive the inflationary potential from a hidden sector Lagrangian. The operator-level identification $V_0\propto M_U^p\,M_P^{4-p}$ is an assumption, and the matching factor $c_p$ is treated as an $\mathcal{O}(1)$ parameter whose impact we quantify explicitly. The question we address is more limited: given the two empirically fixed energy scales, is a single power-law interpolation consistent with $\mathcal{O}(1)$ matching factors, or does it require large corrections that would signal a breakdown?

For the benchmark operator scaling $V_0\propto M_U^4$ with $c_4=3$ (derived from the slow-roll Friedmann relation), the matching requires $\gamma\simeq 0.491$ (derived from endpoint matching) and yields $\beta\simeq 0.68$ (derived). Both values are genuinely $\mathcal{O}(1)$, requiring no large endpoint correction. The required $\gamma$ falls within the range $[0.2,1.0]$ reported for anomalous dimensions in near-conformal gauge theories~\cite{DeGrand2015,Hasenfratz2017}. We also quantify the sensitivity to operator scaling, threshold effects, and observational errors, and verify compatibility with reheating, isocurvature, and $\Delta N_{\rm eff}$ bounds under standard assumptions. A secondary consequence is a late-time curvature threshold of order $|\Omega_K|\sim\mathcal{O}(10^{-4})$, qualitatively distinct from $\Lambda$CDM.

The remainder of this article is organized as follows. Section~\ref{sec:framework} introduces the DRG framework and establishes the late-time anchor. Section~\ref{sec:inflation} specifies the inflationary target. Section~\ref{sec:unified} presents the cross-epoch consistency test. Section~\ref{sec:crosssector} examines cosmological consistency conditions. Section~\ref{sec:future} discusses the late-time curvature consequence. Section~\ref{sec:discussion} addresses interpretation and limitations, and Section~\ref{sec:conclusions} concludes.


\section{Framework and late-time anchor}
\label{sec:framework}

This section introduces the density-responsive gravity (DRG) framework and establishes the late-time anchor for the cross-epoch test. We begin with a physical overview (Section~\ref{sec:DRG_overview}), then specify the effective running mass scale (Section~\ref{sec:RG_scale}), its connection to the density-responsive effective energy (Section~\ref{sec:rho_phi}), and the anchoring procedure (Sections~\ref{sec:beta_anchor} and~\ref{sec:w0_sensitivity}). We use reduced Planck units with $\MP\equiv(8\pi G)^{-1/2}$ throughout. All dimensionful quantities are expressed in units of $\MP$ but we retain $\MP$ explicitly in all equations for clarity.

Cosmological inputs (all observational anchors) are taken from Planck 2018~\cite{Planck2018_cosmo}: $H_0 = 67.4\;\mathrm{km\,s^{-1}\,Mpc^{-1}} = 5.91\times 10^{-61}\,\MP$, $\Omm = 0.315$, $\OmL = 0.685$, giving $\rho_{m,0} = 3\,\Omm\,H_0^2\,\MP^2 \simeq 3.3\times 10^{-121}\,\MP^4$. For the dark-energy equation of state we consider the benchmark value $w_0=-0.989$ (derived from the density-responsive scalar-field closure relations for the fiducial coupling $A=0.024$) and scan over the range $w_0\in[-0.995,-0.98]$, consistent with current constraints from Planck, DES, and DESI~\cite{DES_Y3_2022,DESI2024}.

\subsection{Physical overview of density-responsive gravity}
\label{sec:DRG_overview}

The density-responsive gravity (DRG) framework~\cite{Drobczyk2025DRG} is built on a scalar field $\Phi$ whose effective potential depends explicitly on the ambient matter density $X\equiv\rho_m$. At the equilibrium field value, the scalar contributes an effective gravitating energy density that enters the background expansion (derived in Section~\ref{sec:rho_phi}),
\begin{equation}
\rho_\Phi(X,\mu)
= A\,\MU^4(\mu)\,\frac{1+2s}{(1+s)^2}\,,
\qquad s\equiv\frac{X}{\MU^4(\mu)}\,,
\label{eq:rho_phi_overview}
\end{equation}
Equation~\eqref{eq:rho_phi_overview} interpolates between two cosmologically relevant regimes. At late times ($s\ll 1$), $\rho_\Phi\simeq A\,\MU^4$: a nearly constant vacuum energy that drives cosmic acceleration with equation of state $w=-(1+s)/(1+2s)\simeq -1+\mathcal{O}(s)$. At high ambient densities ($s\gg 1$), $\rho_\Phi\propto\MU^8/X$: the scalar contribution is dynamically suppressed, providing a built-in screening mechanism that evades solar-system and laboratory fifth-force constraints by many orders of magnitude (see~\cite{Drobczyk2025DRG} for quantitative bounds).

The same mass scale $\MU$ enters other applications of the DRG framework~\cite{Drobczyk2025SIDM}, but these play no role in the present analysis, which requires only the late-time regime and the assumption that a standard slow-roll inflationary phase with Hubble scale $H_*$ is realized.

The connecting thread is the effective scaling of $\MU(\mu)$: at late times $\MU(H_0)$ sets the dark-energy density. At inflationary energies $\MU(H_*)$ controls the plateau height. The present study asks whether a single power-law scaling can consistently connect these two anchors.

\subsection{Effective running mass scale and normalization}
\label{sec:RG_scale}

The central dynamical object is a characteristic mass scale $\MU$ whose value depends on the renormalization scale $\mu$ through a power law (assumed ansatz),
\begin{equation}
\MU(\mu) = \beta^{1/4}\,\MP\left(\frac{\mu}{\MP}\right)^\gamma,
\label{eq:MU_def}
\end{equation}
equivalently
\begin{equation}
\MU^4(\mu) = \beta\,\MP^4
\left(\frac{\mu}{\MP}\right)^{4\gamma}.
\label{eq:MU4_beta}
\end{equation}
The effective exponent $\gamma$ encodes the scaling behavior. In strongly coupled SU($N$) gauge theories near the conformal window, anomalous dimensions of fermion bilinears can reach $\gamma_m\sim\mathcal{O}(0.1$--$1)$ depending on the theory, scheme, and operator definition~\cite{Appelquist1988,Sannino2004,DeGrand2015,Hasenfratz2017}. We adopt $\gamma\simeq 0.5$ as a representative starting value (benchmark assumption) and explore robustness over $\gamma\in[0.45,0.55]$.

The dimensionless constant $\beta$ encapsulates threshold corrections, scheme dependence, and possible frame-transformation factors. It cannot be computed without a complete UV theory. However, as we show below, $\beta$ can be anchored using late-time cosmological data, converting it from a free parameter into a derived quantity.

Throughout this study we identify the renormalization scale with the Hubble parameter,
\begin{equation}
\mu = H\,,
\label{eq:mu_H}
\end{equation}
motivated by the role of $H$ as the infrared cutoff for quantum fluctuations in quasi-de~Sitter backgrounds~\cite{Tsamis_Woodard1996,Boyanovsky2006}. Alternative identifications ($\mu\propto\sqrt{\rho}$, $\mu\propto\sqrt{R}$, or $\mu=\kappa H$ with $\kappa\neq 1$) are discussed in \ref{app:scale_setting}. Moderate variations shift $\beta$ but do not alter the main conclusions.

\subsection{Density-responsive effective energy density}
\label{sec:rho_phi}

In the density-responsive gravity (DRG) framework of \cite{Drobczyk2025DRG}, a scalar field $\Phi$ couples to the ambient matter density $X\equiv\rho_m$ through an effective potential. Before specifying the form of this potential, we summarize the cosmological function of $\Phi$, since the present analysis depends on this identification.

\paragraph{Cosmological role of $\Phi$.}
$\Phi$ mediates the density-dependent dark-energy sector of DRG. Its field value is dynamically tied to the local matter density through a minimum of an effective potential $U(\Phi,\rho_m)$~\cite{Drobczyk2025DRG}. Because the relaxation time of $\Phi$ to its instantaneous minimum is short on cosmological scales, $\Phi$ does not act as a slowly rolling quintessence field with its own kinetic dark-energy density. Instead, the gravitating energy of the configuration takes the tracking-vacuum form derived below. At high matter density ($s\!\gg\!1$, deep in the matter-dominated era) it is subdominant and tracks $\rho_\Phi\!\propto\!\rho_m^{-1}$. At low matter density ($s\!\ll\!1$, late times) it asymptotes to a slowly varying vacuum-like plateau $\rho_\Phi\!\simeq\!A\,\MU^4$. In this sense $\Phi$ acts as an effective tracking dark-energy component whose vacuum contribution is set by the running characteristic scale $\MU(\mu)$ rather than by an explicit cosmological constant. This is the distinguishing structural feature of the DRG framework relative to quintessence and chameleon-screened theories. $\Phi$ is not a freely evolving inflaton-like field, and its cosmological imprint enters through the late-time density-dependent vacuum amplitude rather than through a kinetic dark-energy term.

At the equilibrium field value, the minimum of the effective potential contributes
\begin{equation}
U_{\min}(X,\mu) = \frac{A\,\MU^8(\mu)}{X+\MU^4(\mu)}
= A\,\MU^4(\mu)\,\frac{1}{1+s}\,,
\qquad s\equiv\frac{X}{\MU^4(\mu)}\,.
\label{eq:Umin}
\end{equation}
However, because the equilibrium value depends explicitly on the ambient density $X$, the quantity entering the Friedmann equation is the gravitating energy density, obtained from the thermodynamic identity appropriate for a density-dependent minimum,\footnote{In \cite{Drobczyk2025DRG} this step is derived from the full stress-energy tensor. The result is equivalent to the Legendre-type subtraction shown here.}
\begin{equation}
\rho_\Phi(X,\mu)
\equiv U_{\min} - X\,\frac{\partial U_{\min}}{\partial X}
= A\,\MU^4(\mu)\,\frac{1+2s}{(1+s)^2}\,,
\label{eq:rho_phi}
\end{equation}
with corresponding pressure $p_\Phi = -U_{\min}$ and equation of state
\begin{equation}
w(s) \equiv \frac{p_\Phi}{\rho_\Phi}
= -\,\frac{1+s}{1+2s}\,.
\label{eq:w_of_s}
\end{equation}
The pressure remains $p_\Phi=-U_{\min}$ because the minimum contributes as a potential term to the scalar stress energy, whereas the extra term in $\rho_\Phi$ arises from the explicit $X$-dependence of the equilibrium configuration (see \cite{Drobczyk2025DRG} for the full $T_{\mu\nu}$ derivation). Physically, $\rho_\Phi$ interpolates between a tracking regime ($\rho_\Phi\propto \MU^8/\rho_m$ for $s\gg 1$, subdominant at high density) and an approximately constant vacuum-like regime ($\rho_\Phi\simeq A\,\MU^4$ for $s\ll 1$).

\paragraph{Why $\rho_\Phi$ dilutes with cosmic expansion.}
A natural question is why $\rho_\Phi$ is not strictly constant despite arising from the minimum of the effective potential. The resolution is that $U_{\min}$ depends explicitly on the local matter density $X\!=\!\rho_m$ through Equation~\eqref{eq:Umin}. As the universe expands and $\rho_m$ dilutes as $a^{-3}$, the location of the equilibrium configuration of $\Phi$ shifts adiabatically along the family of minima parameterized by $X$. The gravitating energy density is not, therefore, the bare value of $U_{\min}$ at a fixed $\Phi$-configuration (which would be a constant cosmological constant), but the Legendre-subtracted combination Equation~\eqref{eq:rho_phi} that accounts for the $X$-dependence of the equilibrium itself. In the high-density regime one has $\rho_\Phi \simeq 2A\,\MU^8/\rho_m$, so the dilution law is controlled both by the explicit $1/\rho_m$ factor and by the running $\MU(H)\propto H^\gamma$. It is, therefore, not the scaling of a perfect fluid with a fixed equation of state. During an approximately matter-dominated interval, where $H\propto a^{-3/2}$, this gives
\begin{equation}
\rho_\Phi \propto a^{\,3-12\gamma}\qquad (s\gg 1),
\end{equation}
which for the benchmark $\gamma\simeq 0.491$ yields $\rho_\Phi\propto a^{-2.9}$, decreasing rapidly toward the past and remaining subdominant at high density. In the opposite vacuum-like regime ($s\ll 1$) one recovers $\rho_\Phi\simeq A\,\MU^4(H)$, and over an extended matter-like interval this reduces to $\rho_\Phi\propto a^{-6\gamma}$ as used in Section~\ref{sec:future}. The bounded equation-of-state interpolation $w(s)\in[-1,-1/2]$ still captures the qualitative transition from a tracking regime to a vacuum-like regime, but the actual dilution law is set by the density-dependent equilibrium together with the running of $\MU(H)$. This mechanism allows $\Phi$ to act as an effective tracking dark-energy component without reproducing a constant-$\Lambda$ vacuum, and it distinguishes DRG from canonical quintessence, in which the dilution is driven by kinetic energy rather than by a density-dependent equilibrium configuration.

Evaluating at $z=0$ with $s_0\equiv\rho_{m,0}/\MU^4(H_0)$ and inverting $w(s_0)=w_0$ yields the closure relations
\begin{equation}
s_0 = -\frac{1+w_0}{1+2w_0}\,,\qquad
\MU^4(H_0)=\frac{\rho_{m,0}}{s_0}\,,\qquad
A = \frac{\OmL}{\Omm}\,\frac{s_0(1+s_0)^2}{1+2s_0}\,,
\label{eq:late_time_anchor}
\end{equation}
where $\rho_{m,0}=3\,\Omm\,H_0^2\,\MP^2$. These are derived in detail in \ref{app:late_time_anchor}.

Throughout this study we use two complementary perspectives on the late-time parameters. In the \emph{fiducial mode}, the coupling $A$ is fixed from the earlier DRG analysis~\cite{Drobczyk2025DRG} and the closure relations~\eqref{eq:late_time_anchor} are inverted to obtain $w_0$, $s_0$, and $\MU^4(H_0)$. This yields the benchmark $w_0=-0.989$ for $A=0.024$. In the \emph{sensitivity mode} (Section~\ref{sec:w0_sensitivity}), we scan over $w_0$ within observational bounds and compute $A(w_0)$ from the same closure relations. Both modes are algebraically equivalent. The distinction clarifies which quantity is treated as input in each context.

\subsection{Late-time anchoring of $\beta$}
\label{sec:beta_anchor}

Evaluating equation~\eqref{eq:MU4_beta} at $\mu=H_0$ and equating to $\MU^4(H_0)=\rho_{m,0}/s_0$ from equation~\eqref{eq:late_time_anchor} fixes the normalization:
\begin{equation}
\beta = \frac{\MU^4(H_0)}{\MP^4\,(H_0/\MP)^{4\gamma}}\,.
\label{eq:beta_general}
\end{equation}
For $\gamma=1/2$ this simplifies to
\begin{equation}
\beta = \frac{\MU^4(H_0)}{H_0^2\,\MP^2}
= \frac{3\Omm}{s_0}\,.
\label{eq:beta_simple}
\end{equation}

We adopt $A=0.024$ as our fiducial benchmark (benchmark assumption, consistent with~\cite{Drobczyk2025DRG}). Equation~\eqref{eq:beta_general} fixes the normalization only as a function of $\gamma$. Late-time data determine the combination $\beta\,(H_0/\MP)^{4\gamma}$ (equivalently $\MU^4(H_0)$) but do not separately fix $\beta$ and $\gamma$. A reference value used in the walking literature is $\gamma=1/2$, for which the benchmark $w_0=-0.989$ implies $\beta_{1/2}\equiv\beta(0.5)\simeq 84$ (Table~\ref{tab:latetime}). In Section~\ref{sec:unified}, cross-epoch matching provides a second condition that selects $\gamma$ and thereby also fixes $\beta(\gamma)$. The late-time parameters for this reference case are listed in Table~\ref{tab:latetime}.

\begin{table}[htbp]
\centering
\begin{tabular}{lcc}
\hline
Quantity & Value & Reference \\
\hline
$w_0$ & $-0.989$ & implied by $A=0.024$ via equation~\eqref{eq:late_time_anchor} \\
$s_0$ & $0.0112$ & equation~\eqref{eq:late_time_anchor} \\
$\MU^4(H_0)$ & $2.9\times 10^{-119}\,\MP^4$ & $=\rho_{m,0}/s_0$ \\
$\beta_{1/2}\equiv\beta(\gamma\!=\!0.5)$ & $84$ & equation~\eqref{eq:beta_simple} \\
$\beta(\gamma_{\rm best}=0.49135)$ & $0.68$ & Section~\ref{sec:gamma_best} \\
$\OmL^{\rm pred}$ & $0.68$ & closure check \\
\hline
\end{tabular}
\caption{Late-time parameters for the \cite{Drobczyk2025DRG} benchmark $A=0.024$, using Planck 2018 cosmology ($\Omm=0.315$, $\OmL=0.685$, $H_0=67.4\,\mathrm{km\,s^{-1}\,Mpc^{-1}}$). The closure relations~\eqref{eq:late_time_anchor} reproduce $\OmL^{\rm pred}\simeq 0.68$ by construction.\label{tab:latetime}}
\end{table}

For any chosen $\gamma$, Equation~\eqref{eq:beta_general} fixes $\beta(\gamma)$ as a derived quantity, not a free parameter. At the reference $\gamma=0.5$, the value $\beta_{1/2}=84$ may appear large. However, $\beta$ depends sensitively on $\gamma$ and drops to $\mathcal{O}(1)$ near the cross-epoch best fit (Table~\ref{tab:Cinfl_gamma}).\footnote{The log-window assessment and threshold parametrization via $\Xi$ are discussed in Sections~\ref{sec:nature_constraint} and~\ref{sec:gamma_beta_interpretation}.}

The only remaining freedom when extrapolating to inflationary scales is captured by the matching factor $\Cinfl$, which we evaluate in Section~\ref{sec:unified}.

\subsection{Sensitivity to $w_0$}
\label{sec:w0_sensitivity}

Because the inflationary test hinges on $\MU^4(H_0)$ and hence on $\beta$, it is important to quantify how these depend on $w_0$. Table~\ref{tab:w0_scan} shows that $\beta$ varies by a factor of about 4 over $w_0\in[-0.995,-0.980]$. This propagates directly to $V_0^{\rm RG}$ via $\MU^4(H_0)$ and hence to $\Cinfl$. The sensitivity is explored further in the two-dimensional scan of Section~\ref{sec:unified}.

\begin{table}[htbp]
\centering
\begin{tabular}{ccccc}
\hline
$w_0$ & $s_0$ & $A$ & $\beta$ & $\OmL^{\rm pred}$ \\
\hline
$-0.995$ & $0.0051$ & $0.011$ & $187$ & $0.68$ \\
$-0.990$ & $0.0102$ & $0.022$ & $93$  & $0.68$ \\
$-0.989$ & $0.0112$ & $\mathbf{0.024}$ & $\mathbf{84}$ & $0.68$ \\
$-0.985$ & $0.0155$ & $0.034$ & $61$  & $0.67$ \\
$-0.980$ & $0.0208$ & $0.045$ & $45$  & $0.67$ \\
\hline
\end{tabular}
\caption{Sensitivity of late-time parameters to $w_0$ (evaluated at the reference $\gamma=0.5$, i.e.\ $\beta=\beta_{1/2}$ under Planck 2018 cosmology). The predicted $\OmL$ remains close to $0.685$ by construction.\label{tab:w0_scan}}
\end{table}

The recent DES~Y6 combined analysis reports $w_0 = -0.981^{+0.021}_{-0.022}$ with $\Omega_m = 0.305 \pm 0.005$~\cite{DESY6_2024}, which is consistent with the range explored above. Adopting these values shifts $\gamma_{\rm best}$ by $+0.001$ and $\beta(\gamma_{\rm best})$ from $0.68$ to $0.72$, both well within the uncertainty budget of \ref{app:uncertainty_budget}.

The benchmark value $w_0=-0.989$ is not chosen ad~hoc. It is the value implied (derived) by the fiducial coupling $A=0.024$ (benchmark assumption) through the closure relation~\eqref{eq:late_time_anchor} under Planck~2018 cosmology (observational anchor). Reference~\cite{Drobczyk2025DRG} used a slightly different fiducial $A$, yielding the rounder value $w_0\simeq -0.99$. The difference is within rounding, and we use the self-consistent pair $(A=0.024,\;w_0=-0.989)$ throughout.


\section{Inflationary sector}
\label{sec:inflation}

This section specifies the inflationary potential, computes the CMB-normalized plateau height $V_0^{A_s}$ (observational anchor), and establishes the target for the cross-epoch consistency test described in Section~\ref{sec:unified}.

\subsection{Starobinsky potential and slow roll}
\label{sec:potential}

We adopt a Starobinsky-like plateau potential in the Einstein frame~\cite{Starobinsky1980,Whitt1984},
\begin{equation}
V(\phi) = V_0\bigl(1 - e^{-\sqrt{2/3}\,\phi/\MP}\bigr)^2
\equiv V_0\,(1-u)^2,
\label{eq:V_starobinsky}
\end{equation}
where $u\equiv e^{-\sqrt{2/3}\,\phi/\MP}$. This potential arises from $R+R^2$ gravity after conformal transformation and belongs to the broader class of $\alpha$-attractor models~\cite{Kallosh2013,Galante2015}. For $\phi\gg\MP$ the potential reaches a plateau at $V_0$. Inflation ends when $\epsilon_V\equiv(\MP^2/2)(V'/V)^2=1$, giving $\phi_{\rm end}\simeq 0.94\,\MP$.

In principle, a running normalization $V_0(\mu(\phi))$ could modify the potential shape. For $\gamma\simeq 0.5$, however, such prescriptions either produce an unacceptably large tensor-to-scalar ratio or introduce a formal singularity at $\gamma=1/2$. We therefore keep the Starobinsky shape fixed and use the effective scaling only to predict the overall plateau height (see \ref{app:RG_shape} for details).

For the CMB pivot scale $k_*=0.05\,\mathrm{Mpc}^{-1}$ we adopt $N_*=55$ e-folds as our fiducial value, yielding $\phi_*\simeq 5.35\,\MP$ and slow-roll parameters $\epsilon_*\simeq 2.2\times 10^{-4}$, $\eta_*\simeq -1.7\times 10^{-2}$. The resulting spectral observables,
\begin{equation}
n_s \simeq 0.965\,,\qquad r\simeq 0.0035\,,
\label{eq:ns_r}
\end{equation}
lie at the Planck 2018 best-fit value for $n_s$~\cite{Planck2018_inflation} and well below the BICEP/Keck bound $r<0.036$~\cite{BK18}, within reach of LiteBIRD ($\sigma_r\sim 0.001$)~\cite{LiteBIRD2023} and CMB-S4~\cite{CMBS4_2016}. Varying $N_*$ over $[50,60]$ shifts $n_s$ by $\pm 0.003$ and $r$ by $\sim 30\%$, both of which are subdominant to the current uncertainties.

\subsection{CMB normalization and $V_0^{A_s}$}
\label{sec:amplitude}

The scalar power spectrum amplitude $A_s=(2.10\pm 0.03)\times 10^{-9}$~\cite{Planck2018_inflation} fixes the potential at horizon crossing via
\begin{equation}
V_* = 24\pi^2\,\MP^4\,\epsilon_*\,A_s
\simeq 1.09\times 10^{-10}\,\MP^4.
\label{eq:Vstar}
\end{equation}
The asymptotic plateau height follows from $V_*=V_0(1-u_*)^2$ with $(1-u_*)^2\simeq 0.975$:
\begin{equation}
V_0^{A_s} \simeq 1.12\times 10^{-10}\,\MP^4.
\label{eq:V0_As}
\end{equation}
The corresponding inflationary Hubble scale is
\begin{equation}
H_* = \sqrt{\frac{V_*}{3\MP^2}}
\simeq 6.0\times 10^{-6}\,\MP
\simeq 1.5\times 10^{13}\,\mathrm{GeV}.
\label{eq:Hstar}
\end{equation}
Table~\ref{tab:inflation_summary} lists the key inflationary parameters and the dependence on $N_*$ is shown for completeness.\footnote{Over $N_*\in[50,60]$, the CMB-normalized plateau height varies as $V_0^{A_s}\in[0.94,\,1.34]\times 10^{-10}\,\MP^4$ and the best-fit $\gamma$ shifts by $\delta\gamma\simeq 0.001$, subdominant to the viable-band width (Section~\ref{sec:unified}).}

\begin{table}[htbp]
\centering
\begin{tabular}{lc}
\hline
Quantity & Value \\
\hline
$n_s$ & $0.965\;[0.962,\,0.968]$ \\
$r$ & $0.0035\;[0.0030,\,0.0042]$ \\
$V_0^{A_s}/\MP^4$ & $1.12\times 10^{-10}\;[0.94,\,1.34]\times 10^{-10}$ \\
$H_*/\MP$ & $6.0\times 10^{-6}\;[5.6,\,6.6]\times 10^{-6}$ \\
\hline
\end{tabular}
\caption{Inflationary parameters for the Starobinsky potential. The fiducial case is $N_*=55$. Bracketed values indicate the range over $N_*\in[50,60]$.\label{tab:inflation_summary}}
\end{table}

Equation~\eqref{eq:V0_As} defines the observational target for the cross-epoch consistency test.


\section{Cross-epoch consistency test}
\label{sec:unified}

We now ask whether the same effective scaling, anchored by late-time cosmology, can account for the inflationary plateau height. We treat the inflationary phase as an effective description. The consistency requirement derived below does not depend on the pre-inflationary history as long as a standard slow-roll epoch with Hubble scale $H_*$ is realized.

\subsection{A universal endpoint-consistency filter}
\label{sec:universal_filter}

Before specializing to DRG, the core algebra can be written in a model-independent form. Consider any late-time framework that supplies a benchmark anchor scale $M_0 \equiv M(H_0)$ and assume that the same scale obeys the effective power law
\begin{equation}
M(\mu)=\beta^{1/4}\,\MP\left(\frac{\mu}{\MP}\right)^\gamma
\label{eq:M_general_running}
\end{equation}
between $H_0$ and $H_*$. If the inflationary plateau is related to this scale by
\begin{equation}
V_0^{\rm pred}=c_p\,\Xi\,M^p(H_*)\,\MP^{4-p},
\label{eq:V0_general_endpoint}
\end{equation}
where $c_p$ is an $\mathcal{O}(1)$ endpoint factor and $\Xi$ collects cumulative threshold or scheme corrections, then eliminating $\beta$ in favor of the late-time anchor gives
\begin{equation}
V_0^{\rm pred}
= c_p\,\Xi\,M_0^p\,\MP^{4-p}
\left(\frac{H_*}{H_0}\right)^{p\gamma}.
\label{eq:V0_endpoint_filter}
\end{equation}
Matching to the CMB-normalized plateau height $V_0^{A_s}$ yields the general endpoint-consistency relation
\begin{equation}
\gamma
= \frac{1}{p\,\ln(H_*/H_0)}
\ln\!\left(
\frac{V_0^{A_s}}
{c_p\,\Xi\,M_0^p\,\MP^{4-p}}
\right).
\label{eq:gamma_endpoint_filter}
\end{equation}
The large hierarchy $H_*/H_0\sim 10^{55}$ therefore acts as a sharp endpoint filter. Small shifts in the endpoint inputs propagate as
\begin{equation}
\delta\gamma \simeq
\frac{\delta\ln V_0^{A_s}
      - p\,\delta\ln M_0
      - \delta\ln c_p
      - \delta\ln \Xi}
     {p\,\ln(H_*/H_0)}\,.
\label{eq:dgamma_endpoint_filter}
\end{equation}
For the benchmark case $p=4$, one decade of additional endpoint matching, $|\delta\ln X|=\ln 10$, shifts the inferred exponent by only $\delta\gamma\simeq 0.0045$. This hierarchy-enhanced endpoint stiffness is not specific to DRG. DRG enters below only through the explicit identification $M_0\rightarrow \MU(H_0)$ supplied by the density-responsive scalar-field closure relations.

\subsection{DRG benchmark: predicted plateau height and $\Cinfl$}
\label{sec:Cinfl}

Evaluating the effective scaling~\eqref{eq:MU4_beta} at $\mu=H_*$ and substituting the late-time anchor for $\beta$ (Equation~\eqref{eq:beta_general}) yields a manifestly $\beta$-free prediction for the inflationary plateau height:
\begin{equation}
V_0^{\rm RG} = c_4\,\MU^4(H_0)
\left(\frac{H_*}{H_0}\right)^{4\gamma},
\label{eq:V0_RG}
\end{equation}
where $c_4$ is an endpoint matching factor (benchmark assumption) encoding the mapping between the hidden sector scale $\MU$ and the inflationary plateau. We adopt $c_4=3$, motivated by the slow-roll Friedmann relation $V_*\simeq 3\MP^2 H_*^2$ at horizon crossing. The Starobinsky potential relates plateau and pivot via $V_0 = V_*/(1-u_*)^2$ with $(1-u_*)^2\simeq 0.975$ for $N_*=55$, so $V_0\simeq 1.026\,V_*\simeq 3.1\,\MP^2 H_*^2$. Rounding to $c_4=3$ absorbs this $\mathcal{O}(3\%)$ correction, well within the log-window criterion applied to $\Cinfl$. We treat $c_4$ as an $\mathcal{O}(1)$ parameter whose impact is quantified through the log-window criterion and the generalized $p$-scaling analysis (Section~\ref{sec:operator_scaling}). The residual matching factor is
\begin{equation}
\Cinfl \equiv \frac{V_0^{A_s}}{V_0^{\rm RG}}\,.
\label{eq:Cinfl_def}
\end{equation}
The framework requires $\Cinfl=\mathcal{O}(1)$: a value $\Cinfl\gg 1$ or $\ll 1$ would signal a breakdown of the single-power-law connection or the need for additional physics.

Because the lever arm $H_*/H_0\sim 10^{55}$ is enormous, $V_0^{\rm RG}$ is exponentially sensitive to $\gamma$:
\begin{equation}
\frac{\delta V_0^{\rm RG}}{V_0^{\rm RG}}
= 4\ln\!\left(\frac{H_*}{H_0}\right)\delta\gamma
\simeq 506\,\delta\gamma\,.
\label{eq:sensitivity}
\end{equation}
A shift $\delta\gamma=0.01$ changes $V_0^{\rm RG}$ by a factor $\sim 150$. This sensitivity makes $\Cinfl$ a sharp test: given two empirically fixed scales and a single power-law ansatz, the exponent is determined algebraically. For the benchmark $(p,c_4)=(4,3)$ and $\Cinfl=1$,
\begin{equation}
4\gamma_{\rm best}
= \frac{\ln\!\bigl(V_0^{A_s}/
  [3\,\MU^4(H_0)]\bigr)}{\ln(H_*/H_0)}
\simeq 1.965\,,
\label{eq:4gamma_benchmark}
\end{equation}
i.e.\ $\gamma_{\rm best}\simeq 0.491$. The detailed viable band and its dependence on $p$, $c_p$, and $\Xi$ are quantified next.

\subsection{Best-fit $\gamma$ and viable band}
\label{sec:gamma_best}

Setting $\Cinfl=1$ in equations~\eqref{eq:V0_RG}--\eqref{eq:Cinfl_def} and solving for $\gamma$:
\begin{equation}
\gamma = \frac{1}{4}\,
\frac{\ln\bigl(V_0^{A_s}/[c_4\,\MU^4(H_0)]\bigr)}
     {\ln(H_*/H_0)}\,.
\label{eq:gamma_solution}
\end{equation}
With the benchmark values from tables~\ref{tab:latetime} and~\ref{tab:inflation_summary}:
\begin{equation}
\Cinfl = 1 \quad\Longrightarrow\quad
\gamma_{\rm best} = 0.49135\,,\qquad
\beta(\gamma_{\rm best}) = 0.68\,.
\label{eq:gamma_best}
\end{equation}
We round to $\gamma\simeq 0.491$ in the following discussion.

Table~\ref{tab:Cinfl_gamma} lists $\Cinfl$ as a function of $\gamma$. The viable window $\Cinfl\in[0.3,3]$ corresponds to $\gamma\in[0.489,\,0.494]$ with width $\Delta\gamma\simeq 0.005$. The one-decade window $\Cinfl\in[0.1,10]$ yields $\gamma\in[0.487,\,0.496]$ (see Section~\ref{sec:nature_constraint} for the log-window criterion). For comparison, evaluating at $\gamma=0.50$ yields $\Cinfl\simeq 0.013$, meaning the predicted scale overshoots the CMB target by a factor of about 80.

\begin{table}[htbp]
\centering
\begin{tabular}{cccc}
\hline
$\gamma$ & $\beta(\gamma)$ & $\Cinfl$ & Status \\
\hline
$0.485$ & $0.020$ & $24.9$  & $\gg 1$ \\
$0.489$ & $0.19$  & $3.3$   & marginal \\
$0.490$ & $0.32$  & $2.0$   & viable \\
$0.491$ & $0.56$  & $1.2$   & viable \\
$\mathbf{0.49135}$ & $\mathbf{0.68}$ & $\mathbf{1.0}$
  & \textbf{exact match} \\
$0.492$ & $0.98$  & $0.72$  & viable \\
$0.493$ & $1.7$   & $0.43$  & viable \\
$0.494$ & $3.0$   & $0.26$  & marginal \\
$0.500$ & $83.6$    & $0.013$ & $\ll 1$ \\
\hline
\end{tabular}
\caption{Normalization $\beta(\gamma)$ and matching factor $\Cinfl$ as a function of $\gamma$ for the benchmark late-time anchor ($A=0.024$, $c_4=3$).  At the best fit, both $\beta$ and $\Cinfl$ are $\mathcal{O}(1)$.\label{tab:Cinfl_gamma}}
\end{table}

\begin{figure}[htbp]
\centering
\includegraphics[width=\textwidth]{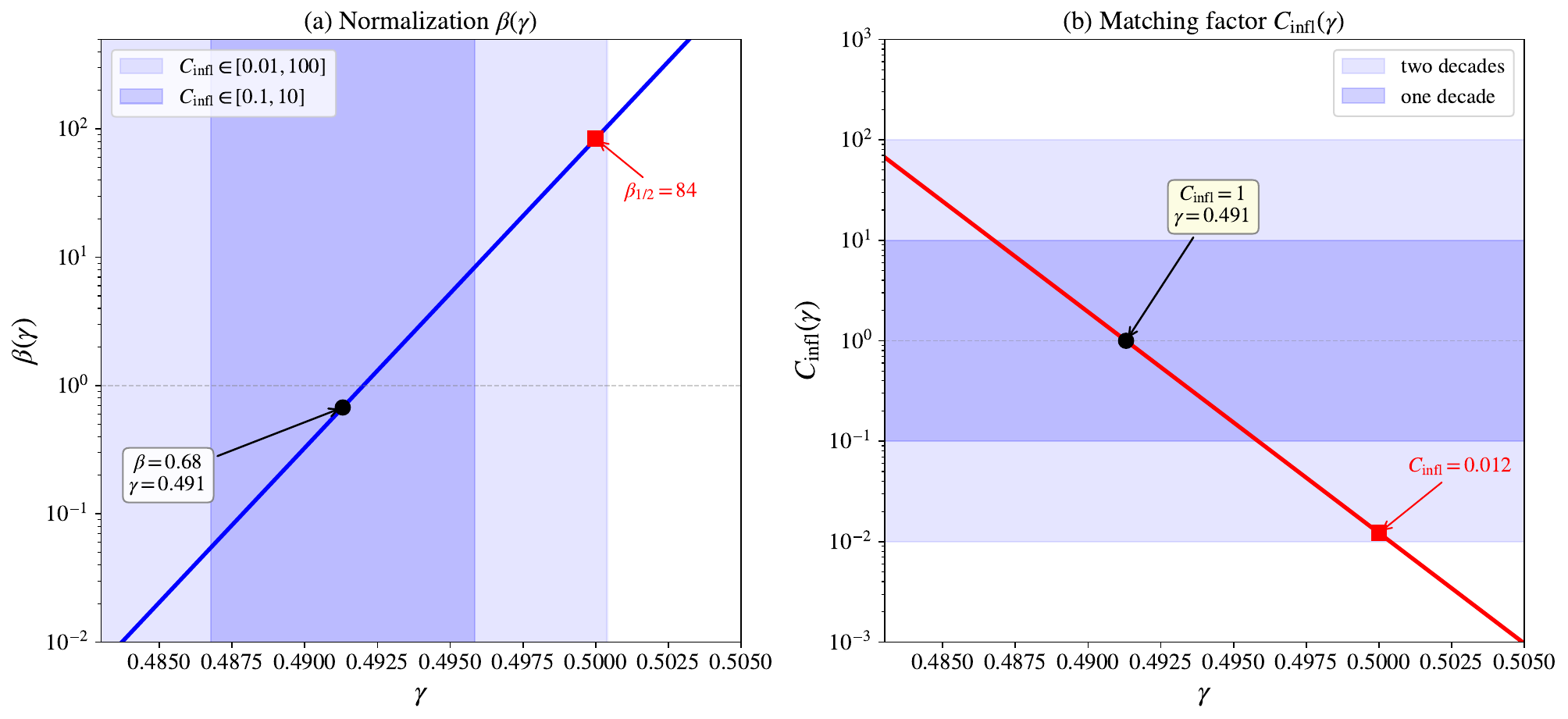}
\caption{Left: normalization $\beta(\gamma)$ from the late-time
anchor. Right: matching factor $\Cinfl(\gamma)$. Shaded bands indicate the one-decade ($\Cinfl\in[0.1,10]$) and two-decade ($\Cinfl\in[0.01,100]$) viable windows. The black circle marks the best fit $\gamma=0.491$. The red square marks the reference $\gamma=0.5$. At the best fit, $\beta\simeq 0.68$ and $\Cinfl\simeq 1$.\label{fig:beta_Cinfl_gamma}}
\end{figure}

Figure~\ref{fig:beta_Cinfl_gamma} shows $\beta(\gamma)$ and $\Cinfl(\gamma)$ across the viable range.

The required $\gamma\simeq 0.491$ is close to but distinguishable from $\gamma=0.50$. The difference admits several interpretations. Walking gauge theories predict $\gamma\sim 0.5$ but not with percent-level precision. Lattice studies report $\gamma_m\in[0.2,1.0]$ depending on $N_f$ and the operator considered~\cite{DeGrand2015,Hasenfratz2017}. Different composite operators can have slightly different scaling, and using $\mu=\kappa H$ with $\kappa\neq 1$ shifts $\gamma_{\rm eff}$ by $\delta\gamma\propto\ln\kappa/\ln(H_*/H_0)\sim 10^{-3}$ (\ref{app:scale_setting}). The constraint should be read as a benchmark for comparison with lattice determinations, not as a first-principles prediction.

\subsection{Sensitivity to operator scaling}
\label{sec:operator_scaling}

The baseline identification $V_0^{\rm RG}=c_4\,\MU^4(H_*)$ (with $c_4=3$) assumes that the inflationary plateau scales as the fourth power of the hidden sector mass (benchmark assumption). More generally, one may consider
\begin{equation}
V_0^{\rm RG} = c_p\,\MU^p(H_*)\,\MP^{4-p}\,,
\label{eq:V0_general_p}
\end{equation}
where $p$ reflects the operator dimension connecting the hidden sector to the inflationary potential and $c_p$ is an $\mathcal{O}(1)$ matching parameter. For $p=4$ the slow-roll relation $V\simeq 3H^2\MP^2$ motivates $c_4\simeq 3$. For general $p$ we treat $c_p$ as capturing operator- and scheme-dependent normalization. The $\Cinfl=1$ condition then becomes
\begin{equation}
p\gamma =
\frac{\ln\!\bigl(V_0^{A_s}/
     [c_p\,(\MU^4(H_0))^{p/4}\,\MP^{4-p}]\bigr)}
     {\ln(H_*/H_0)}\,.
\label{eq:pgamma}
\end{equation}

For the baseline, $p=4$, $c_4=3$ recovers the Friedmann prefactor of \eqref{eq:V0_RG}. Table~\ref{tab:p_sensitivity} shows the required $\gamma$ for several values of $p$. For all $p\in[2,8]$, $\gamma_{\rm best}$ falls within the broad lattice range $\gamma_m\in[0.2,1.0]$, but the viability of the cross-epoch picture depends also on the required normalization $\beta(\gamma)$: Natural endpoint matching ($\beta$ and $\Cinfl$ both $\mathcal{O}(1)$) favors $p\simeq 3$--$4$. The two $p=4$ entries compare canonical normalization ($c_p=1$) with the slow-roll-motivated choice ($c_4=3$). Extreme operator dimensions ($p\leq 2$ or $p\geq 6$) remain algebraically viable but require $|\ln\beta|\gg 1$, signaling either fine-tuning or additional physics not captured by the single power law.

\begin{table}[htbp]
\centering
\begin{tabular}{ccccc}
\hline
$p$ & $\gamma_{\rm best}$ & $p\gamma$ & $\beta(\gamma_{\rm best})$
  & Walking-compatible? \\
\hline
$2$ & $0.448$ & $0.90$ & $2.9\times 10^{-11}$ & marginal ($\beta\ll 1$) \\
$3$ & $0.478$ & $1.44$ & $5.3\times 10^{-4}$  & marginal ($\beta\ll 1$) \\
$4$ (canonical, $c_4=1$) & $0.493$ & $1.97$ & $2.2$      & yes \\
$4$ (benchmark, $c_4=3$)  & $0.491$ & $1.97$ & $0.68$     & yes (benchmark) \\
$6$ & $0.509$ & $3.05$ & $9.6\times 10^{3}$    & marginal ($\beta\gg 1$) \\
$8$ & $0.516$ & $4.13$ & $6.3\times 10^{5}$    & marginal ($\beta\gg 1$) \\
\hline
\end{tabular}
\caption{Required $\gamma$ for $\Cinfl=1$ as a function of operator
scaling $p$ in $V_0\propto c_p\,\MU^p\,\MP^{4-p}$, with $c_p=1$ except where noted. Because the late-time anchor fixes $\MU^4(H_0)$, the anchor contribution $[\MU^4(H_0)]^{p/4}$ depends on $p$, causing $\gamma$ to cluster near $0.5$ for all~$p$ (rather than scaling as $\sim 1.96/p$). Natural endpoint matching ($\beta$ and $\Cinfl$ both $\mathcal{O}(1)$) favors $p\simeq 3$--$4$.\label{tab:p_sensitivity}}
\end{table}
The entries $p=2$ and $p=8$ are labeled ``marginal'' because $\gamma_{\rm best}$ lies near the edges of reported bilinear anomalous dimensions. The relevant operator may differ from a fermion bilinear.

\subsection{Viability map}
\label{sec:viability_map}

\begin{figure}[htbp]
\centering
\includegraphics[width=0.9\columnwidth]{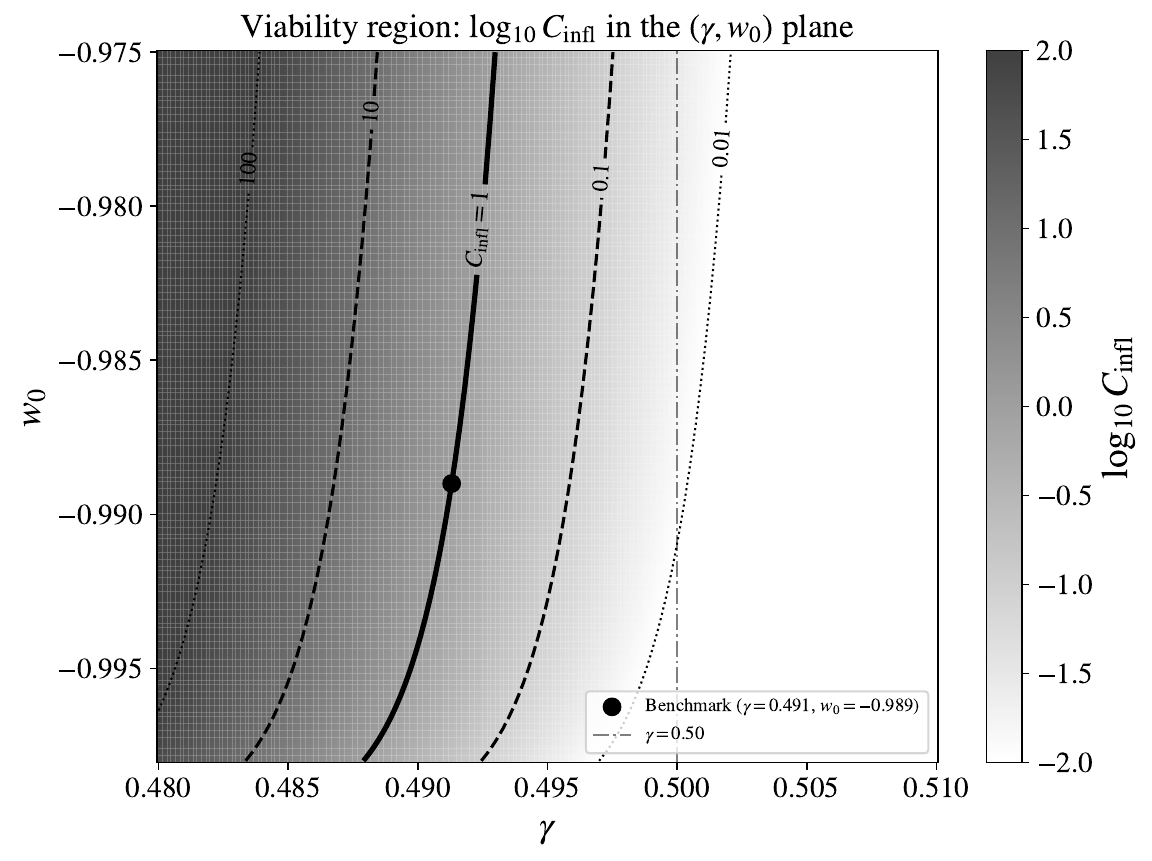}
\caption{Viability region in the $(\gamma,w_0)$ plane for the baseline operator scaling $V_0^{\rm RG}=3\,\MU^4(H_*)$. Solid line: $\Cinfl=1$. Dashed lines: one-decade naturalness window $|\ln \Cinfl|<\ln 10$, i.e.\ $\Cinfl\in[0.1,10]$. Dotted lines: two-decade window $|\ln \Cinfl|<\ln 100$, i.e.\ $\Cinfl\in[0.01,100]$. Circle: benchmark $(\gamma=0.491,\,w_0=-0.989)$. Dash-dotted: $\gamma=0.50$ (naive walking).\label{fig:Cinfl_heatmap}}
\end{figure}

Figure~\ref{fig:Cinfl_heatmap} displays $\log_{10}\Cinfl$ in the $(\gamma,w_0)$ plane for the baseline scaling $p=4$. The $\Cinfl=1$ contour lies at $\gamma\simeq 0.491$, nearly independent of $w_0$. The viable band $\Cinfl\in[0.1,10]$ spans $\gamma\in[0.487,0.496]$ with width $\Delta\gamma\simeq 0.009$. The walking-motivated value $\gamma=0.50$ falls outside the one-decade window but enters the two-decade window $\Cinfl\in[0.01,100]$ at $\gamma\simeq 0.500$.

Physically, a single power law with $\gamma=0.491$ connects \\ $\MU^4(H_0)\sim 10^{-119}\,\MP^4$ (today) to $V_0\sim 10^{-10}\,\MP^4$ (inflation), spanning $\sim 55$ orders of magnitude in $\mu$ and $\sim 110$ in energy density.

Table~\ref{tab:unified_summary} presents the principal results of the cross-epoch test.

\begin{table}[htbp]
\centering
\begin{tabular}{lc}
\hline
Quantity & Value \\
\hline
Late-time anchor $\MU^4(H_0)$ & $2.9\times 10^{-119}\,\MP^4$ \\
CMB normalization $V_0^{A_s}$ & $1.12\times 10^{-10}\,\MP^4$ \\
Scale ratio $H_*/H_0$ & $\sim 10^{55}$ \\
\hline
$\gamma$ for $\Cinfl=1$ & $0.491$ \\
Viable band ($\Cinfl\in[0.1,10]$) & $\gamma\in[0.487,\,0.496]$ \\
$\Cinfl$ at $\gamma=0.50$ & $0.013$ \\
One-decade endpoint shift ($p=4$) & $\delta\gamma\simeq 0.0045$ \\
Stage-IV target $\sigma(w_0)=0.01$ & $\sigma(\gamma)\simeq 0.0018$ \\
\hline
\end{tabular}
\caption{Summary of the cross-epoch consistency test.\label{tab:unified_summary}}
\end{table}

\subsection{Pilot Bayesian extraction of $(\gamma, \beta)$}
\label{sec:pilot_bayesian}

The cross-epoch consistency analysis above is algebraic. Given the inflation-sector anchor $A_s$ (Equation~\eqref{eq:V0_As}) and the late-time DRG amplitude anchor $A=0.024$ (Section~\ref{sec:rho_phi}), the closure relation Equation~\eqref{eq:gamma_endpoint_filter} determines $\gamma$ and Equation~\eqref{eq:beta_general} fixes $\beta(\gamma)$. We present here a complementary Bayesian proof-of-concept in which $\gamma$ and $\log_{10}\beta$ are treated as free parameters of a Markov Chain Monte Carlo (MCMC) sampler, the closure is enforced via two likelihood constraints, and the resulting anchor-prior posterior is reported with its full uncertainty structure. The pilot provides a methodological basis for the recommendation in Section~\ref{sec:observational_prospects} that a full Bayesian extraction be pursued in follow-up work.

\paragraph{Sampler and likelihood.}
We use the affine-invariant ensemble sampler \texttt{emcee}~\cite{ForemanMackey2013} with $64$ walkers and $2\times 10^4$ steps, after a $5\times 10^3$-step burn-in. The sampled parameters are $\boldsymbol{\theta}=(\gamma,\log_{10}\beta,H_0,\Omega_m,\varepsilon_*,A_s)$. The four external parameters $(H_0,\Omega_m,\varepsilon_*,A_s)$ carry gaussian anchor priors, with $(H_0,\Omega_m,A_s)$ observationally anchored and $\varepsilon_*$ assigned a slow-roll width. The numerical anchors are $H_0\!=\!67.66\pm 0.42\,\mathrm{km/s/Mpc}$ (Planck~2018,~\cite{Planck2018_cosmo}), $\Omega_m\!=\!0.3153\pm 0.0073$ (Planck~2018), $\varepsilon_*\!=\!2.2\!\times\!10^{-4}$ with a $30\%$ width reflecting $N_*\in[50,60]$, and $A_s\!=\!(2.099\pm 0.030)\!\times\!10^{-9}$ (Planck~2018~\cite{Planck2018_inflation}). Two likelihood constraints close the system. The late-time amplitude constraint takes the form $A_{\rm pred}(\boldsymbol\theta)\!=\!(\OmL/\Omm)\,s_0(1+s_0)^2/(1+2s_0)$ with $s_0\!=\!\rho_{m,0}/\MU^4(H_0)$ from Equation~\eqref{eq:late_time_anchor}, weighted against the DRG fiducial $A\!=\!0.024\pm 0.001$. The cross-epoch consistency constraint enforces $\log_{10}\!\Cinfl(\boldsymbol\theta)\!\to\!0$ with width $\sigma_{\log_{10}\!\Cinfl}\!=\!0.5$ dex, encoding the Section~\ref{sec:Cinfl} naturalness window $\Cinfl\in[0.1,10]$ at the $\sim\!2\sigma$ level. The forward model uses the closure of Section~\ref{sec:rho_phi} (Equations~\eqref{eq:Umin}--\eqref{eq:late_time_anchor}) and the inflation-sector relation Equation~\eqref{eq:V0_endpoint_filter}. The Planck~2018 values listed above provide the late-time anchor set used here. By construction, the pilot is, therefore, an anchor-prior consistency posterior rather than a full joint fit to the late-time cosmological likelihood.

\begin{figure}[htbp]
\centering
\includegraphics[width=0.95\textwidth]{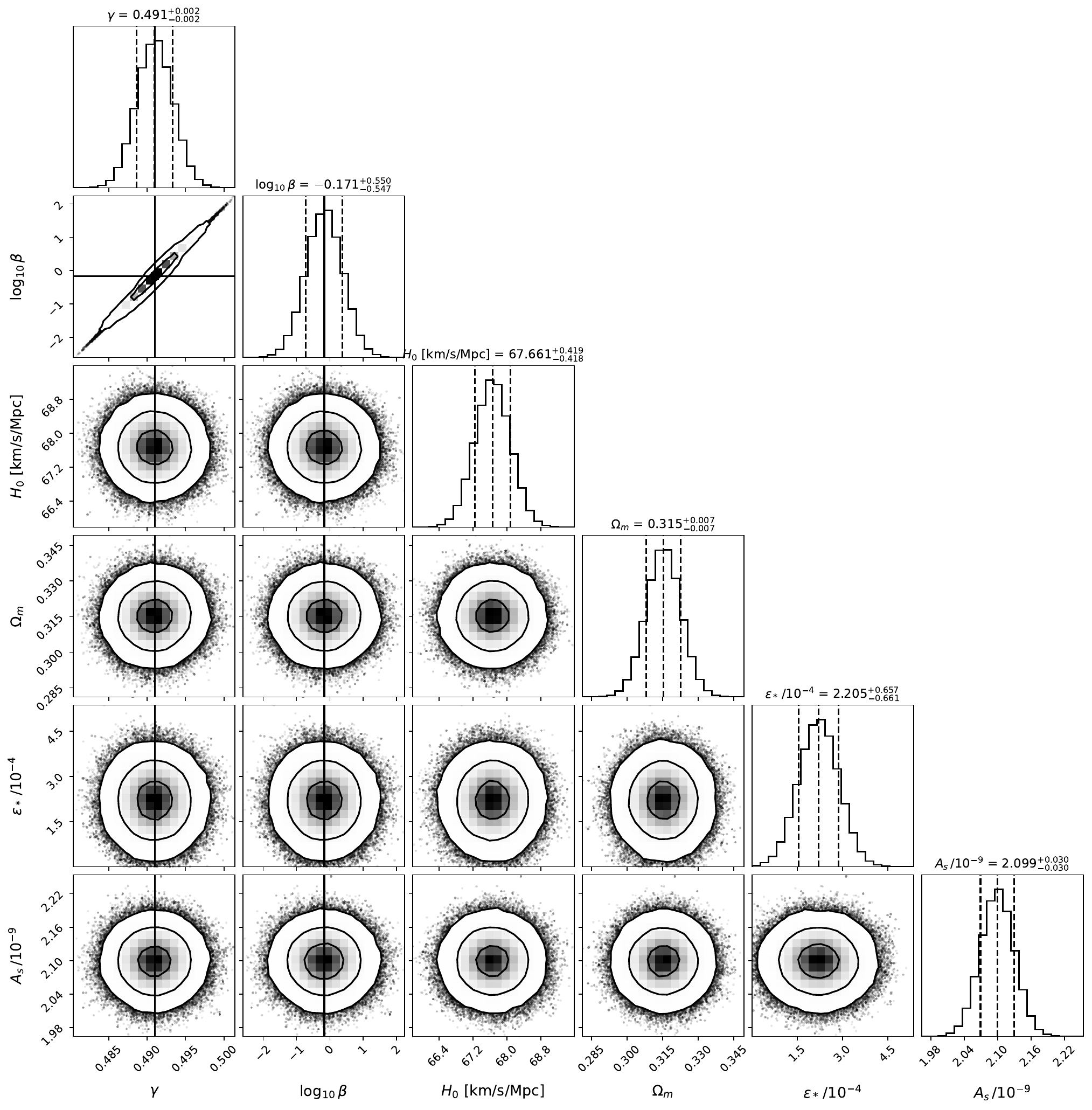}
\caption{Pilot Bayesian posterior on $(\gamma,\log_{10}\beta,H_0,\Omega_m,\varepsilon_*,A_s)$ from a $64\!\times\!2\!\times\!10^4$-step \texttt{emcee} run with Planck~2018 anchors and the two likelihood constraints described in the text. The narrow $\gamma$ posterior (top diagonal) sits at $\gamma\!=\!0.491\pm 0.002$, matching the algebraic best-fit of Section~\ref{sec:gamma_best}. The $(\gamma,\log_{10}\beta)$ panel shows the expected anti-correlation along the late-time degenerate direction. Black solid line: algebraic best-fit $(\gamma,\beta)\!=\!(0.491,0.68)$. Diagonal titles report posterior median $+1\sigma/-1\sigma$. \label{fig:pilot_corner}}
\end{figure}

\paragraph{Posteriors and convergence diagnostics.}
The run converges to
\begin{equation}
\gamma \;=\; 0.4910 \;\pm\; 0.0024,
\qquad
\log_{10}\beta \;=\; -0.171 \;\pm\; 0.55,
\qquad
\beta \;=\; 0.67^{+1.68}_{-0.48}\,.
\label{eq:pilot_results}
\end{equation}
Both posterior medians agree with the algebraic best-fit $(\gamma,\beta)\!=\!(0.491,0.68)$ of Section~\ref{sec:gamma_best} to four significant figures. For the coupled \texttt{emcee} ensemble we diagnose convergence via integrated autocorrelation times rather than Gelman--Rubin. The acceptance fraction is $0.516$, the integrated autocorrelation times are $\tau_{\rm int}\!\simeq\!68$ for both $\gamma$ and $\log_{10}\beta$, and the retained post-burn chain length corresponds to more than $220$ autocorrelation times per walker. This yields a conservative ensemble effective sample count $N_{\rm eff}\!\simeq\!1.4\times 10^4$ for the two parameters of primary interest. The $(\gamma,\log_{10}\beta)$ joint posterior shows the expected anti-correlation along the late-time degenerate direction $\beta\,(H_0/\MP)^{4\gamma}\!=\!\mathrm{const}$ from Equation~\eqref{eq:beta_general}. Figure~\ref{fig:pilot_corner} displays the full $6\times 6$ corner plot of the posterior.

\paragraph{Robustness scan and discussion.}
A sensitivity scan over the cross-epoch tolerance $\sigma_{\log_{10}\!\Cinfl}\in[0.20,1.00]$ dex returns $\gamma$-medians fixed at $0.491$ to four-figure precision across the full scan range, with the posterior width growing linearly from $\sigma(\gamma)\!\simeq\!0.001$ (tight tolerance) to $\sigma(\gamma)\!\simeq\!0.005$ (one-decade tolerance). The $\gamma$-best-fit is, therefore, robust to the precise choice of naturalness window. We emphasize that the pilot is best read as an anchor-prior proof-of-concept rather than as a full cosmological MCMC. The late-time information enters only through gaussian anchor priors on $H_0,\Omega_m,A_s$, the slow-roll prior on $\varepsilon_*$, and the DRG amplitude $A$, not through the full Planck~2018 + DESI~DR2 + Pantheon$+$ + DES-Y3 likelihood that a Cobaya~\cite{Torrado2021}/CAMB~\cite{Lewis1999} implementation would use. A full extraction of this kind is identified as the natural follow-up of the present cross-epoch consistency analysis (cf.\ Section~\ref{sec:observational_prospects}).


\section{Cosmological consistency checks}
\label{sec:crosssector}

Attributing both cosmic accelerations to a hidden sector with effective scaling dynamics imposes additional requirements beyond the amplitude matching of Section~\ref{sec:unified}. Reheating must succeed, isocurvature perturbations must be suppressed, and the hidden sector contribution to $\Neff$ must satisfy CMB bounds. These are necessary conditions that any complete model realization must satisfy. We check them here at the level of order-of-magnitude requirements.

\subsection{Reheating and Big Bang Nucleosynthesis (BBN)}
\label{sec:reheating}

For the Starobinsky potential~\eqref{eq:V_starobinsky}, the inflaton mass at the minimum is $m_\phi=\sqrt{4V_0/3\,\MP^2}\simeq 1.2\times 10^{-5}\,\MP \simeq 3\times 10^{13}\,\mathrm{GeV}$. For Starobinsky inflation, scalaron reheating via gravitational-strength couplings yields
\begin{equation}
T_{\rm RH}\sim 10^{9}\text{--}10^{10}\,\mathrm{GeV}\,,
\label{eq:T_RH}
\end{equation}
depending on the particle content and decay channels~\cite{Bezrukov2009,Gorbunov2011}, well above the BBN threshold $T_{\rm BBN}\simeq 4\,\mathrm{MeV}$ and compatible with gravitino bounds in supersymmetric extensions~\cite{Kawasaki2006}. If hidden sector couplings $g_\chi$ enhance the total decay width, $T_{\rm RH}$ increases further. Reheating does not appear to be a limiting constraint under standard assumptions.

\subsection{Isocurvature perturbations and $\Delta\Neff$}
\label{sec:iso_neff}

During inflation, light fields acquire quantum fluctuations that can source isocurvature modes constrained by Planck to an isocurvature fraction $f_{\rm iso}<0.038$ (95\% CL)~\cite{Planck2018_inflation,Gordon2001}. The key suppression criterion is that hidden sector fields be heavy during inflation:
\begin{equation}
m_{\rm HS}(\phi_*)\gtrsim\mathcal{O}(10)\,H_*\,.
\label{eq:iso_criterion}
\end{equation}
For couplings between the inflaton and the hidden sector of Yukawa or scalar-portal type, this requires $y_\chi\gtrsim 10^{-5}$. This is a mild condition, though the precise requirement is UV model-dependent and can easily yield $m_\chi/H_*\gg 1$.

Hidden sector particles that remain relativistic at BBN or recombination contribute to $\Neff$. Planck constrains $\Delta\Neff<0.30$ (95\% CL)~\cite{Planck2018_cosmo}, which for a hidden sector that thermalizes with the SM and decouples at $T_{\rm dec}$ translates to
\begin{equation}
\Delta\Neff = \frac{g_{\rm HS}}{g_\nu}\,
\left(\frac{T_{\rm HS}}{T_\nu}\right)^4,
\label{eq:Delta_Neff}
\end{equation}
where $g_\nu=7/4$ for a single Weyl neutrino species and $T_\nu/T_\gamma=(4/11)^{1/3}$. For early decoupling ($T_{\rm dec}\gtrsim 100\,\mathrm{GeV}$), adiabatic cooling gives $T_{\rm HS}/T_\nu\simeq (g_{*s}^{\rm dec}/g_{*s}^{\nu\,\rm dec})^{-1/3}\simeq 0.46$, so that $\Delta\Neff\simeq 0.026\,g_{\rm HS}$ and the Planck bound is satisfied for $g_{\rm HS}\lesssim 11$. In walking gauge theories with confinement scale $\Lambda_H\sim\mathrm{TeV}$, the lightest composites (Pseudo-Nambu-Goldstone-Bosons (PNGBs), glueballs) have masses $m\sim\Lambda_H\gg T_{\rm BBN}$ and are non-relativistic at BBN, satisfying the bound automatically. If the hidden sector never thermalizes with the Standard Model (SM), $\Delta\Neff$ is negligible.

Table~\ref{tab:crosssector_summary} summarizes the status of all three checks.

\begin{table}[htbp]
\centering
\begin{tabular}{l l p{6.5cm} l}
\hline
Constraint & Requirement & Sufficient condition & Status \\
\hline
Reheating & $T_{\rm RH}>4\,\mathrm{MeV}$
  & scalaron decay & \checkmark \\[4pt]
Isocurvature & $f_{\rm iso}<0.038$
  & $y_\chi\gtrsim 10^{-5}$ & \checkmark$^*$ \\[4pt]
$\Delta\Neff$ & $\Delta\Neff<0.30$
  & early decoupling ($T_{\rm dec}\gtrsim 100\,\mathrm{GeV}$,
    $g_{\rm HS}\lesssim \mathcal{O}(10)$) or
    non-thermalized/confining HS with
    $m_{\rm light}\gg T_{\rm BBN}$
  & \checkmark \\
\hline
\end{tabular}
\caption{Cosmological consistency checks. All three conditions are satisfied for standard Starobinsky reheating and mild assumptions on the hidden sector spectrum. $^*$Coupling strength $y_\chi$ is model-dependent but the required value is mild.\label{tab:crosssector_summary}}
\end{table}

None of these conditions rules out the framework. They impose requirements on the hidden sector spectrum (i.e. masses and couplings) that restrict model building but are generically satisfied for walking dynamics with $\Lambda_H\sim\mathrm{TeV}$.

Before turning to interpretation, we examine one further consequence of the effective scaling: its implications for the far-future evolution of the universe (Section~\ref{sec:future}).

\section{Late-time curvature consequence}
\label{sec:future}

The effective scaling that connects the inflationary and late-time scales also has consequences for the far-future expansion history. In the DRG framework, the effective vacuum sector does not remain strictly constant. In the vacuum-like regime ($s\ll 1$) one has $\rho_\Phi\simeq A\,\MU^4(H)$, and with $\MU(H)\propto H^\gamma$ this implies $\rho_\Phi(H)\propto H^{4\gamma}$. During an extended interval in which the expansion is approximately matter-dominated ($H\propto a^{-3/2}$), this translates to
\begin{equation}
\rho_\Phi \;\propto\; a^{-6\gamma}
\equiv a^{-n},
\qquad n=6\gamma.
\label{eq:rhoPhi_scaling}
\end{equation}
For the benchmark $\gamma_{\rm best}\simeq 0.491$ one has $n\simeq 2.95$, so $\rho_\Phi$ dilutes faster than spatial curvature ($\propto a^{-2}$). This differs qualitatively from $\Lambda$CDM, where $\rho_\Lambda=\text{const}$ by assumption. In $\Lambda$CDM, the constant vacuum energy eventually exceeds $|\rho_K|\propto a^{-2}$ for any observationally allowed curvature, preventing recollapse. In DRG, by contrast, $\rho_\Phi$ dilutes faster than curvature for $\gamma > 1/3$, so even mild positive curvature can lead to $H=0$ at a finite scale factor.

For $\gamma<\gamma_c\equiv 1/2$ (as in the benchmark), the minimal quasi-equilibrium closure of~\cite{Drobczyk2025DRG} admits a strongly suppressed residual de~Sitter-like attractor. A critical curvature magnitude $|\Omega_{K,\mathrm{crit}}|(\gamma)$ separates universes that reach turnaround from those that approach this suppressed attractor. For the benchmark parameters, dedicated background scans indicate $|\Omega_{K,\mathrm{crit}}|$ is of order $10^{-4}$, well below current observational bounds on $|\Omega_K|$~\cite{Planck2018_cosmo,DESI_DR2_2025}.

The qualitative contrast with $\Lambda$CDM is immediate. In $\Lambda$CDM, turnaround requires $|\Omega_K|\gtrsim\mathcal{O}(1)$, which is observationally excluded. In DRG with $\gamma\gtrsim 0.49$, even $|\Omega_K|\sim 10^{-3}$ can suffice.

This curvature-dependent turnaround constitutes a testable prediction that distinguishes the framework from $\Lambda$CDM. The DRG benchmark predicts $w_0\simeq -0.989>-1$ and a turnaround threshold at $|\Omega_K|\sim\mathcal{O}(10^{-4})$. Whether the universe is closed and whether $|\Omega_K|$ exceeds this threshold are questions that improved curvature measurements from CMB-S4, Euclid, and DESI can address. A detailed threshold analysis, together with the turnaround dynamics and the subsequent contracting phase, is deferred to a companion study.

\section{Discussion}
\label{sec:discussion}

\subsection{Nature of the consistency condition}
\label{sec:nature_constraint}

The requirement $\gamma\simeq 0.491$ (for baseline $p=4$) is algebraically sharp but should be interpreted carefully. It arises because two empirically fixed energy scales, $\MU^4(H_0)$ from dark energy and $V_0^{A_s}$ from the CMB, are connected by a single power law over a large lever arm. The key point is that this stiffness is not specific to DRG. Equation~\eqref{eq:gamma_endpoint_filter} defines a universal endpoint-consistency filter for any framework that provides a late-time anchor scale. The huge hierarchy between $H_0$ and $H_*$ then compresses even order-of-magnitude endpoint uncertainties into percent-level shifts in $\gamma$.

Three aspects are genuinely non-trivial. First, the required $\gamma$ falls within the range $\gamma_m\in[0.2,1.0]$ reported for anomalous dimensions in near-conformal gauge theories. A value $\gamma=3$ or $\gamma=-0.5$ would have ruled out the walking hypothesis outright. Second, $\Cinfl$ remains $\mathcal{O}(1)$, so no large additional threshold corrections are required beyond the power-law extrapolation. Third, the result is stable under $w_0$ variations within current bounds ($\delta\gamma\lesssim 0.003$).

The framework does not reduce the total number of free parameters compared to $\Lambda$CDM + Starobinsky inflation. The DRG relations~\cite{Drobczyk2025DRG} introduce $A$ and $w_0$, the scaling ansatz adds $\gamma$, and $\beta$ is derived from these. The economy, if any, lies in the structural connection between the two acceleration epochs, which is absent in $\Lambda$CDM. Whether this connection reflects a genuine physical mechanism or a parametric coincidence cannot be determined without a concrete UV completion.

To avoid subjective $\mathcal{O}(1)$ language when assessing endpoint matching factors ($\beta$ and $\Cinfl$), we adopt an explicit log-window criterion: a matching factor $X$ is considered natural if $|\ln X|<\Delta$. These log-window thresholds are not physical priors. They serve as a transparent reporting convention for the amount of endpoint matching required, replacing the otherwise ambiguous $\mathcal{O}(1)$ language. For the baseline $p=4$, the one-decade window $\Cinfl\in[0.1,10]$ corresponds to $\gamma\in[0.487,0.496]$ (width $\Delta\gamma\simeq 0.009$). The two-decade window $\Cinfl\in[0.01,100]$ yields $\gamma\in[0.483,0.500]$, which just includes the naive walking value $\gamma=0.50$. At the cross-epoch best fit ($\Cinfl=1$), the inferred normalization is $\beta(\gamma_{\rm best})=0.68$, genuinely $\mathcal{O}(1)$. Equivalently, $\MU(\MP)=\beta^{1/4}\MP\simeq 0.91\,\MP$: the framework matches the two endpoints without invoking a large normalization correction. For comparison, fixing $\gamma=0.5$ yields $\beta_{1/2}=84$ for the same late-time anchor, illustrating the strong $\beta$--$\gamma$ degeneracy along the curve $\beta\,(H_0/\MP)^{4\gamma}=\mathrm{const}$. Detailed $\gamma$ ranges under different criteria are provided in \ref{app:naturalness_details}.

This general filter also provides a useful reference point for running vacuum models. A quadratic $H^2$ running term defines the same effective endpoint index $\gamma=1/2$ on the present map, even though the underlying running-vacuum construction is structurally different from the pure endpoint power law used here. With the benchmark $p=4$ normalization, that reference point gives $\Cinfl\simeq 0.013$, so a single unbroken endpoint scaling at $\gamma=1/2$ would require an additional matching factor of about $80$ to reproduce the CMB plateau. The point is not that running vacuum models are excluded as late-time descriptions, but that the present cross-epoch filter quantifies how much extra endpoint matching such a reference point would need in this inflationary connection.

The identification $\mu=H$ is well motivated during quasi-de~Sitter phases, but between inflation and late-time acceleration the universe passes through radiation domination, matter domination, and multiple particle-physics thresholds. A single unbroken power law over $\sim 55$ decades in $H$ is therefore an assumption. We parametrize departures by writing $\MU^4(H)=\MU^4(H_0)(H/H_0)^{4\gamma}\,\Xi(H)$, where $\Xi=1$ for the pure power law. The factor $\Xi$ absorbs all departures from the idealized single power law, including epoch-dependent shifts in the effective scale identification and particle-physics threshold corrections. Allowing $|\ln\Xi|<\Delta_\Xi$ broadens the inferred $\gamma$ range by
\begin{equation}
\Delta\gamma_\Xi \simeq
\frac{\Delta_\Xi}{4\ln(H_*/H_0)}
\simeq \frac{\Delta_\Xi}{507}\,.
\label{eq:dgamma_Xi}
\end{equation}
In the generalized matching, $\Xi$ shifts the inferred product $p\gamma$ by $\delta(p\gamma)\simeq-\ln\Xi/\ln(H_*/H_0)$. For one-decade threshold uncertainty ($\Delta_\Xi=\ln 10$), $\Delta\gamma_\Xi\simeq 0.005$, comparable to the baseline viable-band width. Threshold effects broaden the allowed window but do not destroy the framework unless cumulative corrections exceed two decades. The full uncertainty budget, combining $\Xi$, $c_p$, and observational errors, is given in \ref{app:uncertainty_budget}.

\paragraph{Confinement as a dominant source of $\Xi$}
The walking motivation for $\gamma\simeq\mathcal{O}(0.5)$ strictly applies only above the hidden sector confinement scale $\Lambda_H$. Below $\Lambda_H$ the microscopic degrees of freedom are confined into composites, and the interpretation in terms of a fermion-bilinear anomalous dimension does not apply directly. For $\Lambda_H\sim\mathrm{TeV}$ the interval $H_0\to\Lambda_H$ spans about 47 decades in $\mu$, while $\Lambda_H\to H_*$ spans only about 8 decades. The majority of the lever arm thus lies in the confined regime. The power law should be read as an effective interpolation between the two endpoint anchors, with the confinement transition expected to provide the dominant physical contribution to the cumulative factor $\Xi$. The consistency test remains meaningful as long as the net matching across $\Lambda_H$ satisfies $|\ln\Xi|\lesssim\ln 10$ to $\ln 100$, the one-to-two-decade window quantified in Equation~\eqref{eq:dgamma_Xi} and \ref{app:uncertainty_budget}. This makes explicit where the single-power-law assumption is most stressed and how it is parametrically bounded.

\subsection{Relation to inflationary puzzles}
\label{sec:puzzles}

In $\Lambda$CDM with Starobinsky inflation, the plateau height $V_0$ is a free parameter fixed by $A_s$. The roughly 120 orders of magnitude separating it from the dark-energy density are unexplained. The present framework reduces this to a single effective exponent $\gamma$ connecting both scales, though it does not explain the value of $\gamma$ itself. In standard treatments, the inflationary scale $V_0^{1/4}\sim 10^{16}\,$GeV and the dark-energy scale $\rho_\Lambda^{1/4}\sim\mathrm{meV}$ are independent inputs. The scaling ansatz attributes the ratio to accumulated running but does not explain why $\gamma$ takes the required value.

The present analysis tests only the energetic coincidence, namely whether a single running scale can reproduce both the dark-energy and inflationary energy densities. It does not test the dynamical identity of the inflaton with a hidden sector field, and the Starobinsky potential is used only as a phenomenological placeholder. The quantitative result also does not rely on DRG being the unique viable late-time framework. DRG is used here as a concrete benchmark because it makes the endpoint test calculable through explicit density-responsive scalar-field closure relations. Alternative late-time frameworks could be analyzed in the same spirit, but they would generally imply different anchor relations and therefore different inferred scaling parameters.
\subsection{Interpretation of $\gamma\simeq 0.49$ and $\beta(\gamma)$}
\label{sec:gamma_beta_interpretation}

The required $\gamma\simeq 0.491$ lies within the lattice range $\gamma_m\in[0.2,1.0]$ for SU($N$) gauge theories near the conformal window~\cite{DeGrand2015,Hasenfratz2017} but is not uniquely selected by the current data. We define $\MU(\mu)$ as the scale-dependent characteristic scale of a hidden sector composite operator $\mathcal{O}$, for example $\MU(\mu)\equiv\langle\mathcal{O}(\mu)\rangle^{1/d_\mathcal{O}}$, so that $\gamma$ parametrizes the effective scaling of this composite. Mapping to standard lattice anomalous dimensions ($\gamma_m$ for fermion bilinears) is model-dependent and requires specifying which operator defines $\MU$ in a concrete gauge theory. Any hidden sector realization implementing the density-responsive framework with $\mu=H$ must produce a composite mass scale whose effective exponent satisfies $\gamma\simeq 0.49$, or the cross-epoch picture fails.

The relation to the value $\gamma\approx 0.50$ deserves clarification. In \cite{Drobczyk2025DRG} the illustrative value $\gamma\approx 0.501$ was obtained under the convention $\MU(\MP)=\MP$, corresponding to $\beta=1$. The present analysis treats $(\gamma,\beta)$ as jointly determined by two conditions: (i)~the late-time anchor $\MU^4(H_0)=\rho_{m,0}/s_0$ and (ii)~the inflationary matching $\Cinfl=1$. Both parametrizations, $(\gamma=0.501,\,\beta=1)$ and $(\gamma=0.491,\,\beta\simeq 0.68)$, reproduce the same physical $\MU^4(H_0)$. They are different points on the degeneracy curve in the $(\gamma,\beta)$ plane. Late-time observations alone constrain only the combination $\beta\,(H_0/\MP)^{4\gamma}$, not $\gamma$ separately. Inflationary matching provides the second condition that breaks this degeneracy, uniquely selecting $\gamma\simeq 0.491$ and $\beta\simeq 0.68$ along the curve.

The reference value $\beta_{1/2}=84$ would indicate that, at $\gamma=0.5$, the naive scaling formula underestimates $\MU^4$ at $\mu=H_0$. In walking gauge theories, condensate enhancement of this magnitude is well documented~\cite{Yamawaki1996,Sannino2009}. However, this is an artifact of evaluating at $\gamma=0.5$ rather than at the best fit: at $\gamma_{\rm best}=0.491$, $\beta\simeq 0.68$ and no large endpoint correction is required. A rigorous assessment of the natural range of $\beta$ at intermediate $\gamma$ requires a concrete hidden sector model.

\subsection{Observational prospects}
\label{sec:observational_prospects}

The benchmark equation of state $w_0\simeq -0.989$ is a prediction of the DRG closure relations for the fiducial choice $A=0.024$, not of the effective scaling ansatz by itself. This benchmark value deviates from $-1$ at the percent level and is consistent with current constraints: DES~Y6 combined data yield $w_0=-0.981^{+0.021}_{-0.022}$ in $w_0$CDM~\cite{DESY6_2024}, and DESI~DR2 analyses combined with CMB and supernova data do not exclude the quintessence regime $w_0>-1$~\cite{DESI_DR2_2025}. Euclid and DESI aim for $\sigma(w_0)\sim 0.01$--$0.02$, which would probe the benchmark branch selected by the density-responsive scalar-field closure. The DES~Y6 combined $1\sigma$ lower bound $w_0=-1.003$ already borders the phantom regime where the present framework does not apply, illustrating the discriminating power of percent-level measurements.

The primary scaling-sensitive output of the present paper is instead the endpoint-selected window in $p\gamma$ and, for the benchmark choice $p=4$, the corresponding narrow range in $\gamma$. In the baseline case the one-decade naturalness window implies $\gamma\in[0.487,0.496]$ around the benchmark $\gamma\simeq 0.491$, equivalently $p\gamma\simeq 1.97$. Within the DRG benchmark, late-time measurements constrain this scaling branch indirectly through the closure relations. Near the benchmark point, $d\gamma/dw_0\simeq 0.18$, implying $\sigma(w_0)=0.02\rightarrow \sigma(\gamma)\simeq 0.0037$ and $\sigma(w_0)=0.01\rightarrow \sigma(\gamma)\simeq 0.0018$, already a sizeable fraction of the one-decade viable band $\Delta\gamma\simeq 0.009$. A full Fisher analysis lies beyond the scope of the present paper, but this simple propagation shows why Stage-IV equation-of-state measurements are directly relevant to the endpoint test.

The same scaling branch also has a distinctive late-time consequence. Because $\rho_\Phi\propto a^{-6\gamma}$, branches with $\gamma>1/3$ dilute faster than curvature. In the benchmark branch this leads to a curvature-driven turnaround threshold of order $|\Omega_{K,\mathrm{crit}}|\sim\mathcal{O}(10^{-4})$, qualitatively distinct from $\Lambda$CDM and in principle accessible to future curvature measurements. The detailed threshold analysis is deferred to a companion study, but the qualitative prediction that mild positive curvature can trigger recollapse is a direct consequence of the scaling branch rather than of the late-time closure alone.

The Starobinsky prediction $r\simeq 0.0035$ lies within the reach of LiteBIRD~\cite{LiteBIRD2023} and CMB-S4~\cite{CMBS4_2016}, although $r$ tests the inflationary potential shape rather than the effective scaling. The present study does not perform a direct full-likelihood data confrontation. A separate companion analysis lies beyond the scope of the present paper and is not required for the benchmark endpoint-consistency test carried out here.

In explicit candidate theories such as near-conformal SU(3) with $N_f\sim 8$--$12$, reported anomalous dimensions cluster around $\gamma_m\sim 0.3$--$0.8$~\cite{DeGrand2015,Hasenfratz2017}. The required $\gamma\simeq 0.49$ lies within this range, but current lattice uncertainties ($\delta\gamma_m\sim 0.3$) are too large for a meaningful test. A concrete comparison requires identifying the composite operator that defines $\MU$ in a specific gauge theory (Section~\ref{sec:gamma_beta_interpretation}).


\section{Conclusions}
\label{sec:conclusions}

We tested whether a single power-law effective scaling, $\MU(\mu)=\beta^{1/4}\,\MP(\mu/\MP)^\gamma$, anchored by late-time dark-energy observations through the density-responsive scalar-field closure relations, can consistently reproduce the CMB-normalized inflationary plateau height. The main novelty is this quantitative endpoint-consistency test itself. It turns a generic unification idea into a falsifiable benchmark problem between two empirically anchored scales and defines a universal endpoint-consistency filter for any framework that supplies a late-time anchor scale.

For the benchmark operator scaling $V_0\propto\MU^4$ with $c_4=3$, endpoint matching requires $\gamma\simeq 0.491$ and yields $\beta\simeq 0.68$, both genuinely $\mathcal{O}(1)$. Under a one-decade log-window criterion, the viable band is $\gamma\in[0.487,0.496]$. Generalizing to $V_0\propto\MU^p\,\MP^{4-p}$ with $p\in[2,8]$, the required $\gamma$ varies only mildly, approximately $0.45$--$0.52$, and natural endpoint matching favors $p\simeq 3$--$4$. For $p=4$, one decade of endpoint mismatch shifts the inferred exponent by only $\delta\gamma\simeq 0.0045$, which makes the filter hierarchy enhanced and sharply constraining despite the logarithmic dependence.

The non-trivial content is that $\gamma\simeq 0.49$ falls within the range reported for anomalous dimensions in near-conformal gauge theories~\cite{DeGrand2015,Hasenfratz2017}, and that $\Cinfl=\mathcal{O}(1)$ without large endpoint corrections. The analysis relies on three explicit assumptions: a single effective power law over the lever arm $H_*/H_0$, scale identification $\mu=H$ in quasi-de~Sitter regimes, and an operator-level scaling $V_0\propto\MU^p\,\MP^{4-p}$ with $\mathcal{O}(1)$ endpoint factors. Within these assumptions, the framework is compatible with Starobinsky reheating, isocurvature suppression, and $\Delta\Neff$ bounds.

Within the DRG benchmark, the equation of state $w_0\simeq -0.989$ provides the late-time anchor rather than an independent observable of the effective scaling itself. Its observational role is nevertheless important because projected Stage-IV precision at the level $\sigma(w_0)\sim 0.01$ maps to $\sigma(\gamma)\sim 0.002$ around the benchmark, enough to probe a substantial fraction of the viable band. The primary scaling-sensitive output is therefore the narrow endpoint-selected window in $\gamma$ and, more generally, in $p\gamma$. The same branch also implies a secondary late-time consequence, namely a curvature-dependent turnaround at $|\Omega_K|\sim\mathcal{O}(10^{-4})$, qualitatively distinct from $\Lambda$CDM. An effective $H^2$ running reference point at $\gamma=0.50$ would sit outside the one-decade window and require an additional correction factor of about $80$. A detailed analysis of the turnaround dynamics and its connection to cyclic cosmology is deferred to a companion study.

The cross-epoch picture has concrete failure modes. It is disfavored if (a) no viable hidden sector realization yields the required effective exponent, (b) cumulative threshold effects correspond to multi-decade endpoint corrections, or (c) future measurements of $w_0$ exclude percent-level deviations from $-1$.

\section*{Declaration of Competing Interest}
The author declares no competing interests.

\section*{Data availability}
The numerical code and data that support the findings of this article are openly available at Zenodo (DOI: 10.5281/zenodo.18621282, see Ref.~\cite{Drobczyk2026_code}).


\appendix


\section{Scale identification}
\label{app:scale_setting}

In the main text we identify the effective running scale with the Hubble parameter, $\mu=H$, motivated by its role as the infrared cutoff for quantum fluctuations in quasi-de~Sitter backgrounds~\cite{Tsamis_Woodard1996,Starobinsky_Yokoyama1994,Boyanovsky2006,Shapiro_Sola2009}. This appendix discusses alternative prescriptions and quantifies their impact on $\beta$ and $\Cinfl$.

\subsection{Alternative scale identifications}
\label{app:alternatives}

\paragraph{Rescaled Hubble: $\mu=\kappa H$}
A dimensionless factor $\kappa\neq 1$ can arise from IR-cutoff conventions, scheme matching for the effective scaling, or threshold effects. Under $\mu\to\kappa\mu$, the normalization transforms as $\beta\to\kappa^{4\gamma}\beta$. For $\gamma\simeq 0.5$ and $\kappa\in[0.5,2]$, this amounts to a factor $\kappa^2\in[0.25,4]$, an $\mathcal{O}(1)$ effect absorbed by the late-time anchoring.

\paragraph{Energy density: $\mu=\rho^{1/4}$}
When $\mu_\rho=(3H^2\MP^2)^{1/4}=3^{1/4}\sqrt{H\MP}$, the ratio $\mu_\rho/\mu_H=3^{1/4}\sqrt{\MP/H}$ varies by $\sim 10^{27}$ between $H_0$ and $H_*$. This strongly epoch-dependent rescaling spoils the single-power-law extrapolation and is therefore disfavored.

\paragraph{Ricci scalar: $\mu=\sqrt{R/12}$}
For flat Friedmann-Lemaître-Robertson-Walker (FLRW), $R=12H^2+6\dot{H}$. During slow roll and late-time acceleration $|\dot{H}|\ll H^2$, so $\mu_R\simeq H$. During matter domination $R=3H^2$, yielding $\mu_R=H/2$. This is effectively a constant-$\kappa$ rescaling and is benign.

\subsection{Impact on $\beta$ and $\Cinfl$}
\label{app:impact}

Table~\ref{tab:scale_impact} summarizes the effect of different prescriptions on $\beta$ for $\gamma=0.5$.

\begin{table}[htbp]
\centering
\begin{tabular}{lccc}
\hline
Identification & $\mu/H$ & $\beta/\beta_{\rm ref}$ & Comment \\
\hline
$\mu=H$ & $1$ & $1$ & baseline \\
$\mu=2\pi H$ & $2\pi$ & $(2\pi)^2\simeq 39$ & common convention \\
$\mu=H/2$ & $0.5$ & $0.25$ & matter-era $\sqrt{R/12}$ \\
$\mu=\rho^{1/4}$ & $\propto\sqrt{\MP/H}$ & epoch-dependent
  & disfavored \\
\hline
\end{tabular}
\caption{Impact of scale identification on the matching factor $\beta$ for $\gamma=0.5$. The reference case is $\mu=H$.\label{tab:scale_impact}}
\end{table}

If the same constant $\kappa$ is used in both epochs, $\Cinfl$ is invariant because the $\beta$-free prediction~\eqref{eq:V0_RG} contains no explicit $\kappa$, so any constant rescaling cancels after anchoring. A genuine effect arises only if the effective scale identification differs between epochs, $\mu_*=\kappa_*H_*$ and $\mu_0=\kappa_0H_0$ with $\kappa_*/\kappa_0\neq 1$. In that case $\Cinfl$ shifts by $(\kappa_*/\kappa_0)^{-4\gamma}$. For $\gamma\simeq 0.49$ and a factor-of-two mismatch this is an $\mathcal{O}(4)$ effect, which is relevant but not catastrophic. The equivalent shift in $\gamma$ is $\delta\gamma=\gamma\ln(\kappa_*/\kappa_0)/\ln(H_*/H_0) \simeq 2.7\times 10^{-3}$ for $\kappa_*/\kappa_0=2$, comparable to the viable-band half-width.


\section{Modified potential shapes}
\label{app:RG_shape}

In the main text the Starobinsky potential shape is held fixed and only the plateau height $V_0$ is set by the hidden sector scaling. This appendix examines what happens if one instead makes the normalization field dependent via a running scale choice $\mu(\phi)$. We stress that these prescriptions correspond to a different, more model-dependent assumption than that adopted above.

Starting from the generic ansatz
\begin{equation}
V(\phi)=V_0\!\bigl(\mu(\phi)\bigr)\,
\bigl(1-e^{-\sqrt{2/3}\,\phi/\MP}\bigr)^2,\qquad
V_0(\mu)=\beta\,\MP^4\left(\frac{\mu}{\MP}\right)^{4\gamma},
\label{eq:AppB_VRG}
\end{equation}
we consider two illustrative scale choices.

\paragraph{Case I: $\mu=H(\phi)$}
Setting $\mu=\sqrt{V/(3\MP^2)}$ makes equation~\eqref{eq:AppB_VRG} implicit in $V$. Solving formally yields $V\propto(1-u)^{2/(1-2\gamma)}$, so the effective exponent of the plateau turn-off changes from $2$ (Starobinsky) to $p_{\rm eff}=2/(1-2\gamma)$. For $\gamma\simeq 0.49$, $p_{\rm eff}\sim\mathcal{O}(10^2)$, an extremely steep turn-off that enhances $\epsilon_V$ by $(1-2\gamma)^{-2}\sim 10^4$ relative to Starobinsky and would generically produce a very large tensor signal. The formal divergence at $\gamma=1/2$ indicates that this prescription is not robust without specifying the full two-sector effective action.

\paragraph{Case II: $\mu=\phi$}
With $\mu=\phi$, the potential on the would-be plateau ($\phi\gg\MP$) becomes a monomial $V\propto\phi^{4\gamma}\simeq\phi^2$. For $N_*=55$ this predicts $n_s\simeq 0.964$ and $r\simeq 0.15$, ruled out by the current CMB upper bounds on $r$.

Table~\ref{tab:RG_shapes} summarizes the outcomes.

\begin{table}[htbp]
\centering
\begin{tabular}{llccl}
\hline
Prescription & \makecell[l]{Effective\\plateau shape} & $n_s$ & $r$ & Status \\
\hline
\makecell[l]{Fixed shape\\(main text)}
  & Starobinsky & ${\sim}0.965$ & ${\sim}0.004$ & viable \\[3pt]
\makecell[l]{$\mu=H(\phi)$,\\self-consistent}
  & steep turn-off & --- & large & disfavored \\[3pt]
$\mu=\phi$
  & $\propto\phi^{4\gamma}$ & ${\sim}0.96$ & ${\sim}0.15$ & ruled out \\
\hline
\end{tabular}
\caption{Impact of running-scale prescriptions on the inflationary
potential shape and observables for $\gamma\simeq 0.49$.\label{tab:RG_shapes}}
\end{table}

The fixed-shape approach adopted in the main text is the conservative and least model-dependent choice for three reasons. First, it separates the well-constrained Starobinsky predictions for $(n_s,r)$ from the hidden sector test of $V_0$. Second, CMB observables probe a narrow energy range ($V_*/V_0\simeq 0.98$ at $N_*=55$), so evaluating the hidden sector scaling at $\mu\simeq H_*$ captures the leading effect without introducing shape deformations. Third, any residual scheme or threshold effects can be absorbed into the matching factor $\Cinfl$. A complete treatment of coupled inflaton--hidden sector running-scale improvement requires a concrete UV completion and is deferred to future work.


\section{Derivation of the late-time anchor}
\label{app:late_time_anchor}

This appendix summarizes the late-time density-responsive scalar-field closure relations used to anchor the scaling normalization at $\mu=H_0$. In the density-responsive gravity framework, the scalar adjusts to an equilibrium determined by the ambient matter density $X\equiv\rho_m$ and the running scale $\MU(\mu)$. The minimum of the effective potential contributes
\begin{equation}
U_{\min}(X,\mu)
= \frac{A\,\MU^8(\mu)}{X+\MU^4(\mu)}
= A\,\MU^4(\mu)\,\frac{1}{1+s}\,,
\quad s\equiv\frac{X}{\MU^4(\mu)}\,.
\label{eq:AppC_Umin}
\end{equation}
The gravitating energy density is obtained from the thermodynamic identity for a density-dependent minimum (cf.\ equation~\eqref{eq:rho_phi}),
\begin{equation}
\rho_\Phi \equiv U_{\min}-X\,\frac{\partial U_{\min}}{\partial X}
= A\,\MU^4(\mu)\,\frac{1+2s}{(1+s)^2}\,,
\quad
w(s)\equiv\frac{p_\Phi}{\rho_\Phi}
= -\,\frac{1+s}{1+2s}\,.
\label{eq:AppC_rho_w}
\end{equation}
Evaluating at $z=0$ and inverting $w(s_0)=w_0$ yields the closure relations used throughout:
\begin{equation}
s_0 = -\frac{1+w_0}{1+2w_0}\,,\quad
\MU^4(H_0)=\frac{\rho_{m,0}}{s_0}\,,\quad
A = \frac{\OmL}{\Omm}\,\frac{s_0(1+s_0)^2}{1+2s_0}\,,
\label{eq:AppC_closure}
\end{equation}
where $\rho_{m,0}=3\,\Omm\,H_0^2\,\MP^2$. The consistency check $\rho_{\Phi,0}=A\,\MU^4(H_0)(1+2s_0)/(1+s_0)^2\simeq 7.0\times 10^{-121}\,\MP^4$ matches the observed $\rho_{\Lambda,0}=3\OmL H_0^2\,\MP^2\simeq 7.2\times 10^{-121}\,\MP^4$ to within rounding.


\section{Naturalness criterion for matching factors}
\label{app:naturalness_details}

We define a matching factor $X$ as natural within a log-window $\Delta$ if $|\ln X|<\Delta$. Using $\ln C_{\rm infl} = \ln V_0^{A_s} - \ln V_0^{\rm RG}$ and $V_0^{\rm RG}\propto (H_*/H_0)^{4\gamma}$, we obtain
\begin{equation}
\delta\gamma = -\frac{\delta\ln C_{\rm infl}}
                     {4\ln(H_*/H_0)}\,.
\label{eq:dgamma_dcinfl}
\end{equation}
Hence a log-window $|\ln C_{\rm infl}|<\Delta$ corresponds to \\ $\gamma\in[\gamma_{\rm best}-\Delta/(4\ln(H_*/H_0)),\; \gamma_{\rm best}+\Delta/(4\ln(H_*/H_0))]$. With $4\ln(H_*/H_0)\simeq 507$, Table~\ref{tab:naturalness} follows directly.

\begin{table}[htbp]
\centering
\begin{tabular}{lccc}
\hline
Criterion & $C_{\rm infl}$ range & $\gamma$ range
  & Width $\Delta\gamma$ \\
\hline
Strict ($\Delta=\ln 3$) & $[0.33,\,3]$ & $[0.489,\,0.494]$
  & $0.005$ \\
One decade ($\Delta=\ln 10$) & $[0.1,\,10]$ & $[0.487,\,0.496]$
  & $0.009$ \\
Two decades ($\Delta=\ln 100$) & $[0.01,\,100]$
  & $[0.483,\,0.500]$ & $0.017$ \\
\hline
\end{tabular}
\caption{Viable $\gamma$ ranges under different naturalness criteria for $C_{\rm infl}$ (baseline $p=4$, benchmark late-time anchor).\label{tab:naturalness}}
\end{table}

Under the two-decade criterion ($\Delta=\ln 100$), the walking-motivated value $\gamma=0.50$ enters the viable window. Under the strict criterion ($\Delta=\ln 3$), the viable band narrows to $\Delta\gamma\simeq 0.005$. Under the one-decade criterion ($\Delta=\ln 10$), $\gamma=0.50$ remains excluded but the window is broad enough to accommodate the dominant theoretical uncertainties ($c_p$, $\Xi$). The appropriate choice of $\Delta$ depends on the expected magnitude of the threshold and matching corrections, which is model-dependent.


\section{Uncertainty budget}
\label{app:uncertainty_budget}

\subsection*{Threshold parametrization}

We parametrize deviations from a strict single power law by $M_U^4(H)=M_U^4(H_0)(H/H_0)^{4\gamma}\,\Xi(H)$, where $\Xi(H)$ encodes cumulative threshold and \\ epoch-dependent effects (cf.\ Section~\ref{sec:nature_constraint} in the main text for motivation and parametric estimates).

\subsection*{Full error budget}

The total uncertainty on the viable $\gamma$ range receives contributions from several sources. In log-differential form:
\begin{equation}
\delta\ln V_0^{\rm RG}
= 4\gamma\,\delta\ln H_*
+ 4\ln\!\left(\frac{H_*}{H_0}\right)\delta\gamma
+ \delta\ln M_U^4(H_0)
+ \delta\ln\Xi
+ \delta\ln c_p\,,
\label{eq:full_uncertainty}
\end{equation}
where $\Xi$ parametrizes cumulative threshold effects (Section~\ref{sec:nature_constraint}). Solving for $\delta\gamma$ at fixed $C_{\rm infl}$ gives $\delta\gamma = -[\gamma/(4\ln(H_*/H_0))]\,\delta\ln H_*$ for the $H_*$ contribution, and analogously for the other terms.

\begin{table}[htbp]
\centering
\begin{tabular}{llc}
\hline
Source & Magnitude & $\delta\gamma$ \\
\hline
$A_s$ (CMB amplitude) & $\delta\ln A_s\simeq 0.01$ & $\sim 2\times 10^{-5}$ \\
$w_0$ (dark energy EOS) & nonlinear, $w_0\in[-0.995,-0.980]$ & $\sim 3\times 10^{-3}$ \\
$H_*$ (inflationary scale) & $\delta\ln H_*\simeq 0.1$ & $\sim 4\times 10^{-4}$ \\
$N_*$ (e-fold number) & $N_*\in[50,60]$ & $\sim 10^{-4}$--$10^{-3}$ \\
$c_p$ (operator prefactor) & $\delta\ln c_p\simeq\ln 10$ & $\sim 5\times 10^{-3}$ \\
$\Xi$ (threshold effects) & $\delta\ln\Xi\simeq\ln 10$ & $\sim 5\times 10^{-3}$ \\
\hline
\textbf{Total (quadrature)} & & $\sim 0.008$ \\
\hline
\end{tabular}
\caption{Uncertainty contributions to $\gamma_{\rm best}$. Each source is mapped to $\delta\gamma$ via equation~\eqref{eq:full_uncertainty} solved for $\delta\gamma$ at fixed $C_{\rm infl}$.\label{tab:uncertainty_budget}}
\end{table}

The $w_0$ contribution cannot be reliably estimated by linear error propagation around $w_0\simeq -1$ because $s_0(w_0)=-(1+w_0)/(1+2w_0)$ is strongly nonlinear in this regime. Therefore, we quote $\delta\gamma$ from the explicit scan in Table~\ref{tab:w0_scan}.

The dominant uncertainties are the operator prefactor $c_p$ and the possible threshold effects $\Xi$, both of which are model-dependent. Observational uncertainties ($A_s$, $w_0$, $H_*$) are subdominant. The quadrature sum is illustrative. The parameters $c_p$ and $\Xi$ could be correlated in explicit UV completions, in which case the effective uncertainty may differ. The total uncertainty $\delta\gamma\sim 0.008$ is comparable to the baseline viable-band width, confirming that the constraint is meaningful but not over-determined.

\section*{References}
\bibliographystyle{elsarticle-num}
\bibliography{paper}

@article{Drobczyk2025DRG,
doi = {10.1088/1361-6382/ae1ac1},
url = {https://doi.org/10.1088/1361-6382/ae1ac1},
year = {2025},
month = {nov},
publisher = {IOP Publishing},
volume = {42},
number = {22},
pages = {225016},
author = {Drobczyk, Martin},
title = {A density-responsive scalar-field framework for singularity regularization and dynamical dark energy},
journal = {Classical and Quantum Gravity}
}

@article{Drobczyk2025SIDM,
doi = {10.1088/1361-6382/ae16f9},
url = {https://doi.org/10.1088/1361-6382/ae16f9},
year = {2025},
month = {nov},
publisher = {IOP Publishing},
volume = {42},
number = {22},
pages = {225006},
author = {Drobczyk, Martin},
title = {Naturally resonant two-mediator model of self-interacting dark matter with decoupled relic abundance},
journal = {Classical and Quantum Gravity}
}

@misc{Drobczyk2026_code,
  author    = {Drobczyk, Martin},
  title     = {Code for: A cross-epoch endpoint-consistency test of a single effective scaling from dark energy to inflation},
  year      = {2026},
  publisher = {Zenodo},
  doi       = {10.5281/zenodo.18621282},
  url       = {https://doi.org/10.5281/zenodo.18621282},
  note      = {Concept DOI; resolves to latest version},
}

@article{Starobinsky1980,
  author = {Starobinsky, A. A.},
  title = "{A new type of isotropic cosmological models without singularity}",
  journal = {Phys. Lett. B},
  volume = {91},
  pages = {99--102},
  year = {1980},
  doi = {10.1016/0370-2693(80)90670-X}
}

@article{Starobinsky2007,
  author = {Starobinsky, A. A.},
  title = "{Disappearing cosmological constant in f(R) gravity}",
  journal = {JETP Lett.},
  volume = {86},
  pages = {157--163},
  year = {2007},
  eprint = {0706.2041},
  archivePrefix = {arXiv}
}

@article{Whitt1984,
  author = {Whitt, Brian},
  title = "{Fourth-order gravity as general relativity plus matter}",
  journal = {Phys. Lett. B},
  volume = {145},
  pages = {176--178},
  year = {1984},
  doi = {10.1016/0370-2693(84)90332-0}
}

@article{Kallosh2013,
  author = {Kallosh, Renata and Linde, Andrei and Roest, Diederik},
  title = "{Superconformal inflationary $\alpha$-attractors}",
  journal = {JHEP},
  volume = {11},
  pages = {198},
  year = {2013},
  eprint = {1311.0472},
  archivePrefix = {arXiv}
}

@article{Galante2015,
  author = {Galante, Mario and Kallosh, Renata and Linde, Andrei and Roest, Diederik},
  title = "{Unity of cosmological inflation attractors}",
  journal = {Phys. Rev. Lett.},
  volume = {114},
  pages = {141302},
  year = {2015},
  eprint = {1412.3797},
  archivePrefix = {arXiv}
}

@article{Bezrukov2009,
  author = {Bezrukov, Fedor and Shaposhnikov, Mikhail},
  title = "{Standard Model Higgs boson mass from inflation: Two loop analysis}",
  journal = {JHEP},
  volume = {07},
  pages = {089},
  year = {2009},
  eprint = {0904.1537},
  archivePrefix = {arXiv}
}

@article{Gordon2001,
  author = {Gordon, Christopher and Wands, David and Bassett, Bruce A. and Maartens, Roy},
  title = "{Adiabatic and entropy perturbations from inflation}",
  journal = {Phys. Rev. D},
  volume = {63},
  pages = {023506},
  year = {2001},
  eprint = {astro-ph/0009131},
  archivePrefix = {arXiv}
}

@article{Peebles_Vilenkin1999,
  author = {Peebles, P. J. E. and Vilenkin, Alexander},
  title = "{Quintessential inflation}",
  journal = {Phys. Rev. D},
  volume = {59},
  pages = {063505},
  year = {1999},
  eprint = {astro-ph/9810509},
  archivePrefix = {arXiv}
}

@article{Dimopoulos_Owen2017,
  author = {Dimopoulos, Konstantinos and Owen, Charlotte},
  title = "{Quintessential inflation with $\alpha$-attractors}",
  journal = {JCAP},
  volume = {06},
  pages = {027},
  year = {2017},
  eprint = {1703.00305},
  archivePrefix = {arXiv}
}

@article{Appelquist1988,
  author = {Appelquist, Thomas and Karabali, Dimitra and Wijewardhana, L. C. R.},
  title = "{Chiral hierarchies and the flavor changing neutral current problem in technicolor}",
  journal = {Phys. Rev. Lett.},
  volume = {57},
  pages = {957--960},
  year = {1986},
  doi = {10.1103/PhysRevLett.57.957}
}

@article{Yamawaki1996,
  author = {Yamawaki, Koichi and Bando, Masako and Matumoto, Ken-iti},
  title = "{Scale-invariant hypercolor model and a dilaton}",
  journal = {Phys. Rev. Lett.},
  volume = {56},
  pages = {1335--1338},
  year = {1986},
  doi = {10.1103/PhysRevLett.56.1335}
}

@article{Sannino2004,
  author = {Sannino, Francesco and Tuominen, Kimmo},
  title = "{Orientifold theory dynamics and symmetry breaking}",
  journal = {Phys. Rev. D},
  volume = {71},
  pages = {051901},
  year = {2005},
  eprint = {hep-ph/0405209},
  archivePrefix = {arXiv}
}

@article{Sannino2009,
  author = {Sannino, Francesco},
  title = "{Conformal windows of SP(2N) and SO(N) gauge theories}",
  journal = {Phys. Rev. D},
  volume = {79},
  pages = {096007},
  year = {2009},
  eprint = {0902.3494},
  archivePrefix = {arXiv}
}

@article{DeGrand2015,
  author = {DeGrand, Thomas},
  title = "{Lattice tests of beyond Standard Model dynamics}",
  journal = {Rev. Mod. Phys.},
  volume = {88},
  pages = {015001},
  year = {2016},
  eprint = {1510.05018},
  archivePrefix = {arXiv}
}

@article{Hasenfratz2017,
  author = {Hasenfratz, Anna and Schaich, David},
  title = "{Nonperturbative $\beta$ function of twelve-flavor SU(3) gauge theory}",
  journal = {JHEP},
  volume = {02},
  pages = {132},
  year = {2018},
  eprint = {1610.10004},
  archivePrefix = {arXiv}
}

@article{Shapiro_Sola2009,
  author = {Shapiro, Ilya L. and Sol\`a, Joan},
  title = "{On the possible running of the cosmological ``constant''}",
  journal = {Phys. Lett. B},
  volume = {682},
  pages = {105--113},
  year = {2009},
  eprint = {0910.4925},
  archivePrefix = {arXiv}
}

@article{SolaPeracaula2022,
  author = {Sol\`a Peracaula, Joan},
  title = "{Running vacuum in quantum field theory in curved spacetime: renormalizing $\rho_{\rm vac}$ without $\sim m^4$ terms}",
  journal = {Phil. Trans. R. Soc. A},
  volume = {380},
  pages = {20210182},
  year = {2022},
  eprint = {2203.13757},
  archivePrefix = {arXiv}
}

@article{Nojiri_Odintsov2011,
  author = {Nojiri, Shin'ichi and Odintsov, Sergei D.},
  title = "{Unified cosmic history in modified gravity: From F(R) theory to Lorentz non-invariant models}",
  journal = {Phys. Rept.},
  volume = {505},
  pages = {59--144},
  year = {2011},
  eprint = {1011.0544},
  archivePrefix = {arXiv}
}

@article{Gorbunov2011,
  author = {Gorbunov, Dmitry S. and Panin, Alexey G.},
  title = "{Scalaron the mighty: producing dark matter and baryon asymmetry at reheating}",
  journal = {Phys. Lett. B},
  volume = {700},
  pages = {157--162},
  year = {2011},
  eprint = {1009.2448},
  archivePrefix = {arXiv}
}

@article{Kawasaki2006,
  author = {Kawasaki, M. and Kohri, K. and Moroi, T. and Yotsuyanagi, A.},
  title = "{Big-bang nucleosynthesis and gravitino}",
  journal = {Phys. Rev. D},
  volume = {78},
  pages = {065011},
  year = {2008},
  eprint = {0804.3745},
  archivePrefix = {arXiv}
}

@article{Planck2018_cosmo,
  author = "{Planck Collaboration}",
  title = "{Planck 2018 results. {VI}. Cosmological parameters}",
  journal = {Astron. Astrophys.},
  volume = {641},
  pages = {A6},
  year = {2020},
  eprint = {1807.06209},
  archivePrefix = {arXiv}
}

@article{Planck2018_inflation,
  author = "{Planck Collaboration}",
  title = "{Planck 2018 results. {X}. Constraints on inflation}",
  journal = {Astron. Astrophys.},
  volume = {641},
  pages = {A10},
  year = {2020},
  eprint = {1807.06211},
  archivePrefix = {arXiv}
}

@article{BK18,
  author       = {{BICEP/Keck Collaboration}},
  title        = {Improved Constraints on Primordial Gravitational Waves
                  using {Planck}, {WMAP}, and {BICEP/Keck} Observations
                  through the 2018 Observing Season},
  journal      = {Phys. Rev. Lett.},
  volume       = {127},
  pages        = {151301},
  year         = {2021},
  eprint       = {2110.00483},
  archivePrefix= {arXiv},
  primaryClass = {astro-ph.CO}
}

@article{DESI2024,
  author       = {{DESI Collaboration}},
  title        = {{DESI} 2024 {VI}: Cosmological Constraints from the
                   Measurements of Baryon Acoustic Oscillations},
  journal      = {JCAP},
  volume       = {2025},
  number       = {02},
  pages        = {021},
  year         = {2025},
  doi          = {10.1088/1475-7516/2025/02/021},
  eprint       = {2404.03002},
  archivePrefix = {arXiv}
}

@article{DESI_DR2_2025,
  author       = {{DESI Collaboration} and Abdul-Karim, M. and others},
  title        = {{DESI} {DR2} Results {II}: Measurements of Baryon
                   Acoustic Oscillations and Cosmological Constraints},
  journal      = {Phys.\ Rev.\ D},
  volume       = {112},
  pages        = {083515},
  year         = {2025},
  doi          = {10.1103/PhysRevD.112.083515},
  eprint       = {2503.14738},
}

@article{DES_Y3_2022,
  author = "{DES Collaboration}",
  title = "{Dark Energy Survey Year 3 results: Cosmological constraints from galaxy clustering and weak lensing}",
  journal = {Phys. Rev. D},
  volume = {105},
  pages = {023520},
  year = {2022},
  eprint = {2105.13549},
  archivePrefix = {arXiv}
}

@article{DESY6_2024,
  author = "{DES Collaboration}",
  title = "{Dark Energy Survey Year 6 Results: Cosmological Constraints from Galaxy Clustering and Weak Lensing}",
  journal = {arXiv preprint},
  year = {2026},
  eprint = {2601.14559},
  archivePrefix = {arXiv}
}

@article{LiteBIRD2023,
  author = "{LiteBIRD Collaboration}",
  title = "{Probing cosmic inflation with the LiteBIRD cosmic microwave background polarization survey}",
  journal = {Prog. Theor. Exp. Phys.},
  volume = {2023},
  pages = {042F01},
  year = {2023},
  eprint = {2202.02773},
  archivePrefix = {arXiv}
}

@article{CMBS4_2016,
  author = "{CMB-S4 Collaboration}",
  title = "{CMB-S4 Science Book, First Edition}",
  journal = {arXiv preprint},
  year = {2016},
  eprint = {1610.02743},
  archivePrefix = {arXiv}
}

@article{Tsamis_Woodard1996,
  author = {Tsamis, N. C. and Woodard, R. P.},
  title = "{The quantum gravitational back-reaction on inflation}",
  journal = {Ann. Phys.},
  volume = {253},
  pages = {1--54},
  year = {1997},
  eprint = {hep-ph/9602316},
  archivePrefix = {arXiv}
}

@article{Starobinsky_Yokoyama1994,
  author = {Starobinsky, Alexei A. and Yokoyama, Jun'ichi},
  title = "{Equilibrium state of a self-interacting scalar field in the de Sitter background}",
  journal = {Phys. Rev. D},
  volume = {50},
  pages = {6357--6368},
  year = {1994},
  eprint = {astro-ph/9407016},
  archivePrefix = {arXiv}
}

@article{Boyanovsky2006,
  author = {Boyanovsky, D. and de Vega, H. J. and Sanchez, N. G.},
  title = "{Quantum corrections to slow roll inflation and new scaling of superhorizon fluctuations}",
  journal = {Nucl. Phys. B},
  volume = {747},
  pages = {25--54},
  year = {2006},
  eprint = {astro-ph/0503669},
  archivePrefix = {arXiv}
}

@article{Weinberg1989,
  author  = {Weinberg, Steven},
  title   = {The cosmological constant problem},
  journal = {Rev. Mod. Phys.},
  volume  = {61},
  pages   = {1--23},
  year    = {1989},
  doi     = {10.1103/RevModPhys.61.1}
}

@article{Martin2012,
  author  = {Martin, J\'{e}r\^{o}me},
  title   = {Everything you always wanted to know about the cosmological
             constant problem (but were afraid to ask)},
  journal = {Comptes Rendus Physique},
  volume  = {13},
  pages   = {566--665},
  year    = {2012},
  doi     = {10.1016/j.crhy.2012.04.008}
}

@article{ForemanMackey2013,
  author  = {Foreman-Mackey, Daniel and Hogg, David W. and Lang, Dustin
             and Goodman, Jonathan},
  title   = {emcee: The MCMC Hammer},
  journal = {Publ. Astron. Soc. Pac.},
  volume  = {125},
  pages   = {306--312},
  year    = {2013},
  eprint  = {1202.3665},
  archiveprefix = {arXiv}
}

@article{Torrado2021,
  author  = {Torrado, Jes\'us and Lewis, Antony},
  title   = {Cobaya: code for {B}ayesian analysis of hierarchical
             physical models},
  journal = {JCAP},
  volume  = {05},
  pages   = {057},
  year    = {2021},
  eprint  = {2005.05290},
  archiveprefix = {arXiv}
}

@article{Lewis1999,
  author  = {Lewis, Antony and Challinor, Anthony and Lasenby, Anthony},
  title   = {Efficient computation of {CMB} anisotropies in closed {FRW}
             models},
  journal = {Astrophys. J.},
  volume  = {538},
  pages   = {473--476},
  year    = {2000},
  eprint  = {astro-ph/9911177},
  archiveprefix = {arXiv}
}

\end{document}